\documentclass[11pt]{article}
\usepackage{amsfonts}
\usepackage{amsgen,amsmath,amstext,amsbsy,amsopn,amssymb,subfigure,amsthm}
\usepackage{soul}
\usepackage[dvips]{graphicx}
\usepackage{comment}
\usepackage{enumitem}
\usepackage{natbib}
\usepackage{booktabs}
\usepackage{blkarray}
\usepackage{algorithm}
\usepackage{algpseudocode}
\usepackage{url}
\usepackage{longtable}
\usepackage{mathtools}
\usepackage{authblk}
\usepackage[usenames]{color}
\usepackage{cleveref}

 \allowdisplaybreaks

\DeclarePairedDelimiter\ceil{\lceil}{\rceil}

 \allowdisplaybreaks

\newtheorem{Theorem}{Theorem}

\newtheorem{Lemma}{Lemma}
\newtheorem{Remark}{Remark}

\newtheorem{Proposition}{Proposition}
\newtheorem{Assumption}{Assumption}

\newcommand{\be}{\begin{equation}}
\newcommand{\ee}{\end{equation}}
\newcommand{\bea}{\begin{eqnarray}}
\newcommand{\eea}{\end{eqnarray}}
\newcommand{\beas}{\begin{eqnarray*}}
\newcommand{\eeas}{\end{eqnarray*}}

\newcommand{\cT}{\mathcal{T}}

\newcommand{\V}{{\rm V}}
\newcommand{\Var}{{\rm Var}}

\newcommand{\Cov}{{\rm Cov}}

\newcommand{\rank}{{\rm rank}}
\newcommand{\var}{{\rm var}}

\newcommand{\tr}{{\rm tr}}

\newcommand{\diag}{{\rm diag}}

\newcommand{\cov}{{\rm Cov}}

\newcommand{\argmin}{\mathop{\rm arg\min}}

\allowdisplaybreaks

\makeatletter
\newcommand*{\rom}[1]{\expandafter\@slowromancap\romannumeral #1@}
\makeatother

\allowdisplaybreaks

\begin{document}

\title{Nonparametric covariance estimation for mixed longitudinal studies, with applications in midlife women's health}

\date{(\today)}

\author[1,2]{Anru R. Zhang} 
\author[3]{Kehui Chen}
\affil[1]{Department of Statistics, University of Wisconsin-Madison}
\affil[2]{Department of Biostatistics and Bioinformatics, Duke University}
\affil[3]{Department of Statistics, University of Pittsburgh}
\maketitle
\bigskip

\begin{abstract}
	In mixed longitudinal studies, a group of subjects enter the study at different ages (cross-sectional) and are followed for successive years (longitudinal). In the context of such studies, we consider nonparametric covariance estimation with samples of noisy and partially observed functional trajectories. The proposed algorithm is based on a noniterative sequential-aggregation scheme with only basic matrix operations and closed-form solutions in each step. The good performance of the proposed method is supported by both theory and numerical experiments. We also apply the proposed procedure to a study on the working memory of midlife women, based on data from the Study of Women's Health Across the Nation (SWAN).

\end{abstract}

\section{Introduction}

A mixed longitudinal study is a mixture of a longitudinal and a cross-sectional study \citep{berger1986comparison, helms1992intentionally}. Suppose the researchers intend to study the social and cognitive development of children aged four to twelve. In an ideal longitudinal design, a group of four-year-old children will be recruited and followed over eight successive years. In a mixed longitudinal design, one can recruit a group of children between the ages of four and eight, and then follow them for four years (within a typical funding period). Because the age requirement is more flexible at recruitment, this type of mixed longitudinal design results in shorter completion times and potentially larger group sizes.  
However, this type of mixed longitudinal design also brings new challenges for statistical analysis, because the trajectory is only partially observed for each subject.

\begin{figure}[ht]
	\centerline{
		\includegraphics[scale=0.32]{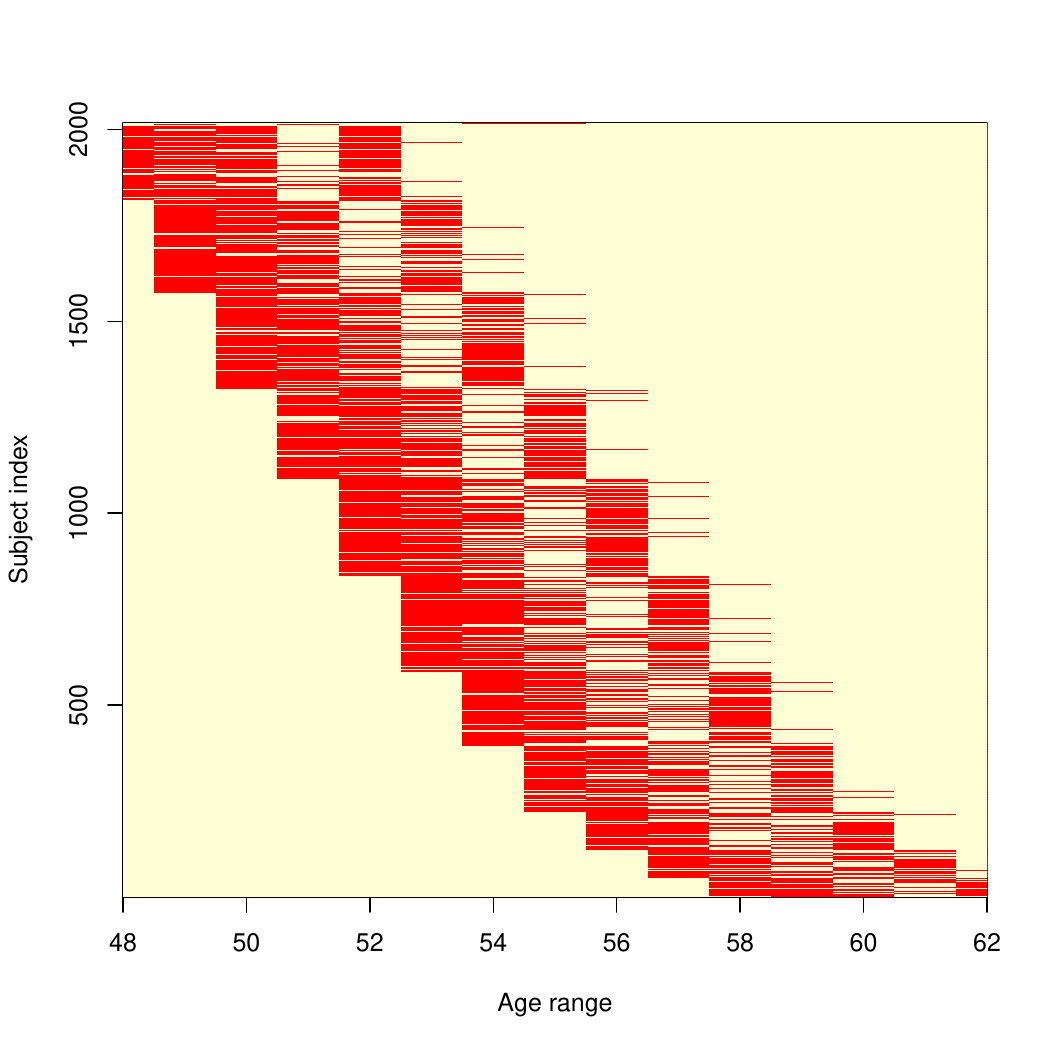}\includegraphics[scale = 0.32]{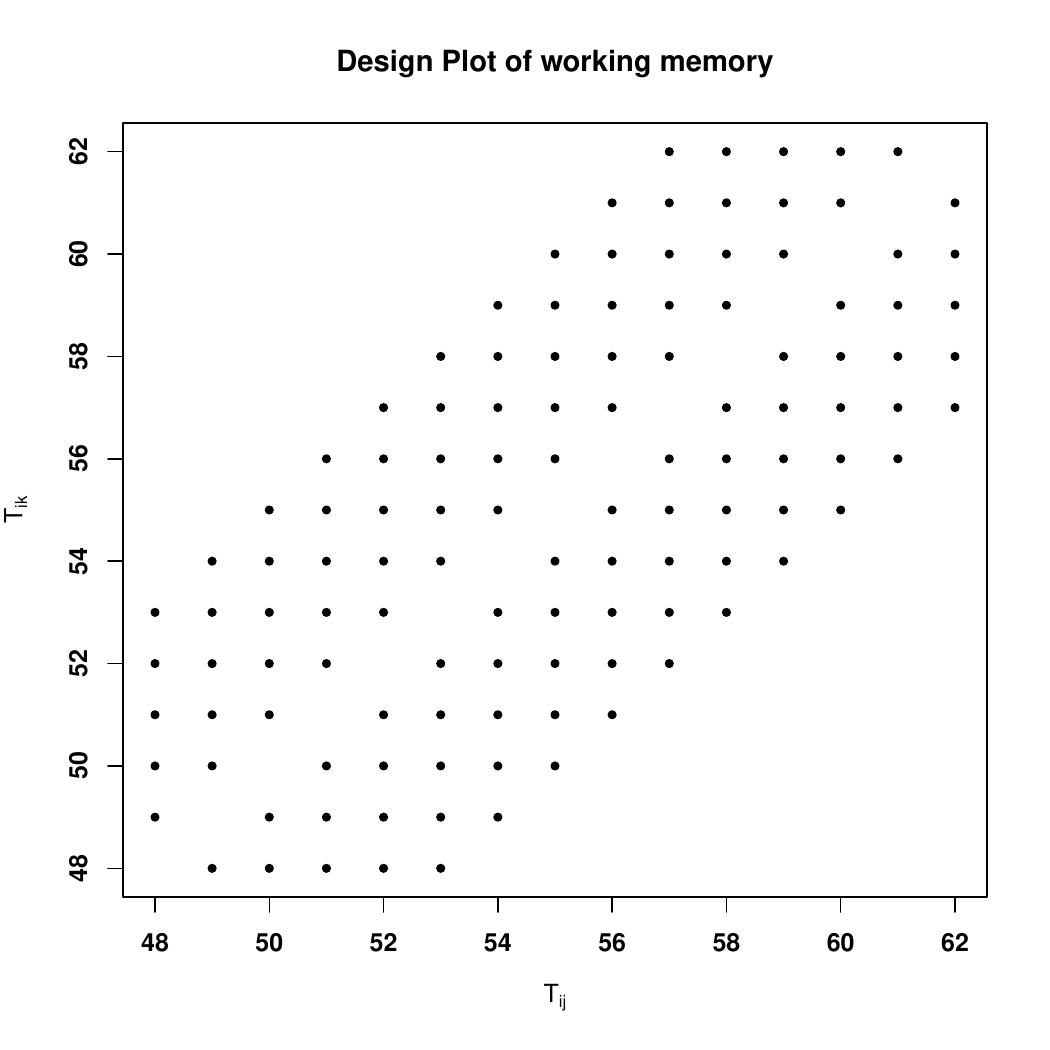}	}
	\caption{Left: For each of the 2016 subjects, measurements were between age $x$ and $x+5$, for some $x \in [48, 58]$. Right: The design plots for covariance $G(s,t)$, that is, the assembled pairs of $(t_{ij}, t_{il})$, for $1\leq i\leq n, 1\leq j <l \leq n_i$. The pooled pairs do not fill the entire domain $\mathcal{T}^2$ because there are no measurements available for pairs of $(t_{ij},t_{il})$ whenever $|j-l| > 5$.}
	\label{Fig:design}
\end{figure}

Specifically, we consider a data example from the Study of Women's Health Across the Nation (SWAN). The SWAN is a community-based, longitudinal study of midlife women. Women aged between 42 and 52 years were enrolled around 1996/97, and followed annually thereafter. Currently, SWAN data up to the 10th follow up visit are available in a publicly accessible repository managed by the ICPSR, at \url{http://www.icpsr.umich.edu/icpsrweb/ICPSR/series/00253}. Although enormous studies have examined cognitive functioning in midlife, few are longitudinal, and most are based on three or fewer cognition assessments \citep{karlamangla2017evidence}.As a result, there are insufficient studies on within-person longitudinal decline in cognitive performance in those under 60 years of age \citep{hedden2004insights,ronnlund2005stability}. In contrast, the SWAN data contain more follow-ups and a wider age range, providing a good opportunity for a longer-term study of women's midlife health. In particular, we focus on working memory measurements available from Visits six to ten. By pooling the subjects, the age range under consideration is a span of 15 years: $\mathcal{T} = [48, 62]$. However, given its mixed longitudinal design, the longitudinal follow-ups of each subject in the SWAN only capture a piece of the chronological aging trajectory, and the shape might have a complex interaction with age \citep{ronnlund2005stability,fuh2006longitudinal}. As shown in the left panel of Figure \ref{Fig:design}, the measurements for each subject are only a subset within a period of at most five years. Traditional parametric models, such as linear mixed models (with age as a between-subject effect, and the time of follow-up as a within-subject effect), often assume a linear trend over time. However, the individual chronological aging trajectories of working memory might have a complex shape. For example, working memory might improve first and then decline, and the age when working memory starts to decline varies among subjects. Therefore, we believe that nonparametric models, such as a functional principal component analysis, may reveal interesting features. 

We consider a mixed longitudinal design for $n$ subjects, where for each subject $k$, measurements are obtained at times $t_{kj}$, for $k = 1, \dots, n$ and $j = 1, \dots, n_k$. We use the notation 
\begin{equation}\label{datamodel}
X_k(t_{kj}) = Z_k(t_{kj}) + \epsilon_{kj}, \ \  t_{kj}\in \mathcal{T},
\end{equation}
where $\epsilon_{kj}$ are zero mean independent and identically distributed (i.i.d.) measurement errors that are uncorrelated with all other random components and satisfy $\var(\epsilon_{kj}) = \sigma^2$. Here, $Z(t)$, for $t \in \mathcal{T}$, is assumed to be a square-integrable random process with mean and covariance functions $\mu(t)$ and $G(s,t)= \cov(Z(s), Z(t))$. In a mixed longitudinal design, the observed time points $\{t_{kj}\}_{j = 1,\dots, n_k}$ for each subject $k$ are restricted to a subject-specific partial domain. As shown in the right panel of Figure \ref{Fig:design}, we do not have within-subject correlation information for any two points that are more than five years apart in the SWAN data example. To apply a functional data approach for mixed longitudinal studies, the main methodological challenge is to nonparametrically estimate the covariance structure $G$ of the underlying process. 

Estimating the mean and covariance functions plays a fundamentally important role in a functional data analysis. Useful tools, such as a functional principal component analysis, often rely on a consistent covariance function estimation \citep{yao2005functional, hall2006properties, li2010uniform}. For conventional functional data, where the pooled design (right panel of Figure \ref{Fig:design}) for the covariance is complete, various methods based on kernel smoothing and splines have been proposed (e.g., \cite{rice1991estimating, yao2005functional, peng2009geometric, xiao2013fast}). In a study in which the covariance information is incomplete, \cite{fan2007analysis} considered a semiparametric covariance estimation, where the variance function $G(t,t) =\sigma^2(t)$ is modeled non-parametrically under smoothness conditions, while the off-diagonal correlation structures are assumed to have a parametric form $\rho(s,t,\theta)$. 
However, this problem differs from the banded covariance estimation considered in studies such as \cite{bickel2008regularized}, \cite{cai2010optimal}, \cite{cai2012adaptive}, \cite{cai2016estimating}, and the references therein, because there is no bandable covariance structure in our scenario, and the design pairs are only within a banded area.

We propose estimating the covariance suing a sequential-aggregation scheme (see Section \ref{sec:method}). The proposed algorithm is noniterative, with closed-form solutions and only basic matrix operations (such as matrix multiplication and singular value decomposition (SVD)) in each step. We prove that under moderate conditions (see Section \ref{sec:theory}), the proposed method consistently recovers the nonparametric covariance structure using data within a banded area. A key step of the proposed procedure is solving the orthogonal Procrustes or Wahba problem \citep{wahba1965least}, that is, finding a rotation matrix to best align two sets of points in two different Euclidean coordinate systems. This problem was first motivated by satellite attitude determination, then later applied to many other applications. To theoretically analyze the procedure, we introduce a new error bound for the solution to Wahba problem (Lemma \ref{lm:rotations-Wahba}). In the theoretical analysis, we introduce a series of technical tools on perturbation inequalities of singular subspaces, including Lemmas \ref{lm:algebra-least-singular-value}, \ref{lm:A-B}, \ref{lm:factorization-lemma}, and \ref{lm:Hy-fan-norm}, which may be of independent interest.

Fragmentary functional observations have been studied under other modeling assumptions; see, for example, \cite{delaigle2013classification} and \cite{delaigle2016approximating}.  \cite{descary2017recovering} and \cite{kneip2017optimal} consider covariance estimation and reconstruction from fragmentary functional observations using an optimization framework. The implementations of both works involve iterations. In particular, \cite{descary2017recovering} formulates the problem as a nonconvex optimization that aims to minimize the error within the observable diagonal band under a rank constraint. In contrast, we introduce a novel sequential-aggregation approach that provides explicit solutions and new insights into the covariance estimation problem. We also include numerical comparisons with the method of \cite{descary2017recovering} in the simulation section. In addition, this problem is related to several recent works on high-dimensional covariance estimation with missing values. For example, \cite{loh2012high} and \cite{lounici2014high} consider a linear regression or covariance matrix estimation, where the observations are missing randomly with a fixed rate. In contrast, \cite{kolar2012consistent} and \cite{cai2016minimax} consider a more general setting that allows a nonrandom missing pattern, but still requires that each pair of covariates simultaneously appear in a sufficient number of samples. 
The problem discussed in this paper is distinct from these existing settings, because a large portion of the covariate pairs will never appear in the same sample (such as the pairs between the earlier and latest observations in the longitudinal studies), by the nature of the design. \cite{bishop2014deterministic} studied a similar sequential-aggregation scheme for matrix completion. However, they mainly consider the completion of high-dimensional low-rank positive semidefinite matrices in a deterministic setting, whereas we provide a statistical guarantee for covariance estimation from partially observed noisy functional data.

The rest of this paper is organized as follows. The methodology and algorithm are described in Section \ref{sec:method}, followed by theoretical analyses in Section \ref{sec:theory}. In Section \ref{sec:numerical}, we present a series of numerical experiments, including the application to the SWAN data. Section \ref{sec:discussion} concludes the paper. The proofs are collected in the Supplementary Materials.

\section{Covariance Estimation for Mixed Longitudinal Design}\label{sec:method}

We briefly introduce the notation that will be used throughout the paper. For a matrix $A\in \mathbb{R}^{p_1\times p_2}$ or bivariate function $G$, let $\{\sigma_1(A), \sigma_2(A), \ldots\}$ and $\{\sigma_1(G), \sigma_2(G),\ldots\}$ be singular values in nonincreasing order. We adapt the R syntax to indicate matrices/functions restricted to the subsets of indices/domains: if $A\in \mathbb{R}^{p_1\times p_2}$, and $a\leq b$ and $c\leq d$ are four positive integers, we use $A_{[a:b, c:d]}$ to denote the submatrix of $A$ formed by its $a$th to $b$th rows and $c$th to $d$th columns. Here, ``:" alone represents the entire index set, so $A_{[:, 1:r]}$ and $A_{[a:b, :]}$ represent the first $r$ columns of $A$ and the $\{a,\ldots, b\}$th rows of $A$, respectively; similarly, $G_{[\mathcal{T}_1, \mathcal{T}_2]}$ represents a function $G$ with domain $\mathcal{T}_1\times \mathcal{T}_2$. Let $\mathcal{L}(\mathcal{T})$ be the Lebesgue measure of any domain $\mathcal{T}$. Let $\|A\|_{F}$ and $\|A\|$ be the matrix Frobenius norm and operator norm, respectively: $\|A\|_F = \left(\sum_{i, j}A_{ij}^2\right)^{1/2} = \left(\sum_{i}\sigma_i^2(A)\right)^{1/2}$, $\|A\| = \sigma_{\max}(A)$. Denote $I_{r\times r}$ as the $r$-by-$r$ identity matrix, and $\mathbb{O}_{p, r} = \{V: V^\top V = I_{r\times r}\}$ as the set of all $p$-by-$r$ matrices with orthonormal columns. In particular, the set of all $r$-by-$r$ orthogonal matrices can be denoted as $\mathbb{O}_r = \mathbb{O}_{r,r}$. Denote $\|G\|_{HS} = \left(\iint |G(s_1, s_2)|^2ds_1s_2\right)^{1/2}$ as the Hilbert\textendash Schmidt norm of the bivariate function $G$. Finally, we use $C, C_0, C_1, c, c_0, \ldots$ to represent generic constants, the exact values of which may vary from line to line.

Suppose $\mathcal{T}$ is the entire period of interest. Consider an equally spaced grid of time points $T = \{t_1,\dots,t_p\}$ on the time domain $\mathcal{T}$. In a mixed-longitudinal design, suppose $\mathcal{T}_k$ is the observational period for subject $k$, and we observe $X_k(T_k)$ in the contiguous band of the domain $\mathcal{T}_k$:
$$\mathcal{T}_k \subseteq \mathcal{T},\quad \frac{ \mathcal{L}(\mathcal{T}_k) }{ \mathcal{L}(\mathcal{T}) }= \delta, \quad T_k \subseteq T\cap \mathcal{T}_k =  \left\{t_1, \ldots, t_p\right\} \cap \mathcal{T}_k, \quad  k = 1,\ldots, n.$$
Here, the fraction of observation $\delta$ is assumed to be a constant between zero and one and $T_k$ might not be consecutive, owing to missing values. If $T_k$ is complete with no missing values, the number of observations is $d$, with $\delta = d/p$. Suppose the signal-noise decomposition \eqref{datamodel} holds for each observation: $X_k(t_{kj}) = Z_k(t_{kj}) + \epsilon_{kj}$. Let $\Sigma_0$ denote the $p\times p$ discretized version of covariance $G$, that is, the $(i,j)$th entry of $\Sigma_0$ is equal to $\Cov(Z(t_i), Z(t_j)) = G(t_i, t_j)$. We estimate $G$ using the discretized version $\Sigma_0$. Suppose $G$ has approximate rank $r$. 
Then, we also have $\Sigma_0 \approx AA^\top$, where $A\in \mathbb{R}^{p\times r}$ can be regarded as the factors of $\Sigma_0$. 

\begin{figure}\label{fig:procedure}
	\begin{center}
		\subfigure[Step 1. Construction of $I_l$, $l=1,\ldots, l_{\max}$]{\includegraphics[height=1.5in]{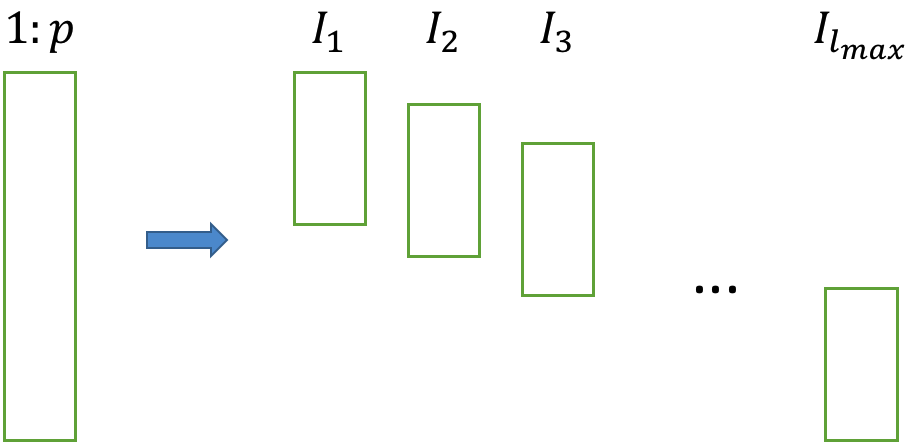}}\\
		\subfigure[Step 2 and 3. Construction of $\hat{\Sigma}_l$ and $\hat{A}_l$, for $l=1,\ldots, l_{\max}$]{\includegraphics[height=1.5in]{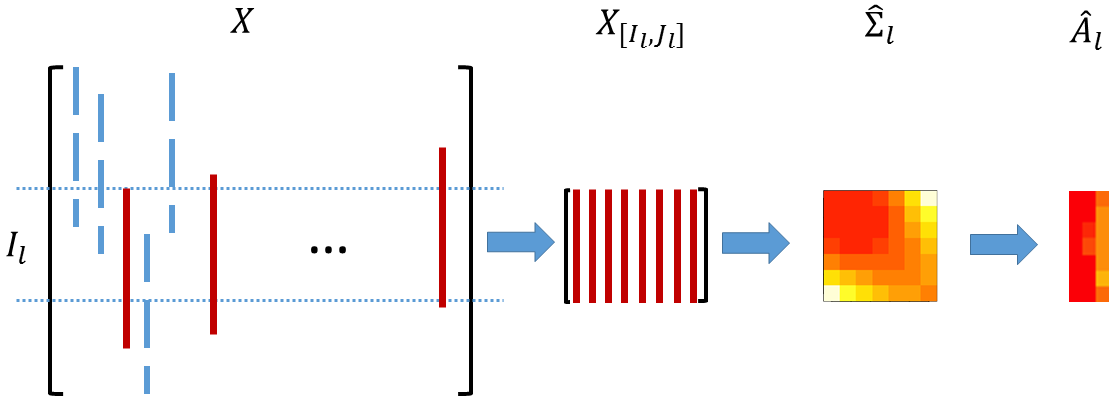}}\\
		\subfigure[Step 4. Rotate $\hat{A}_l$ via $\hat{O}_l$]{\includegraphics[height=1.5in]{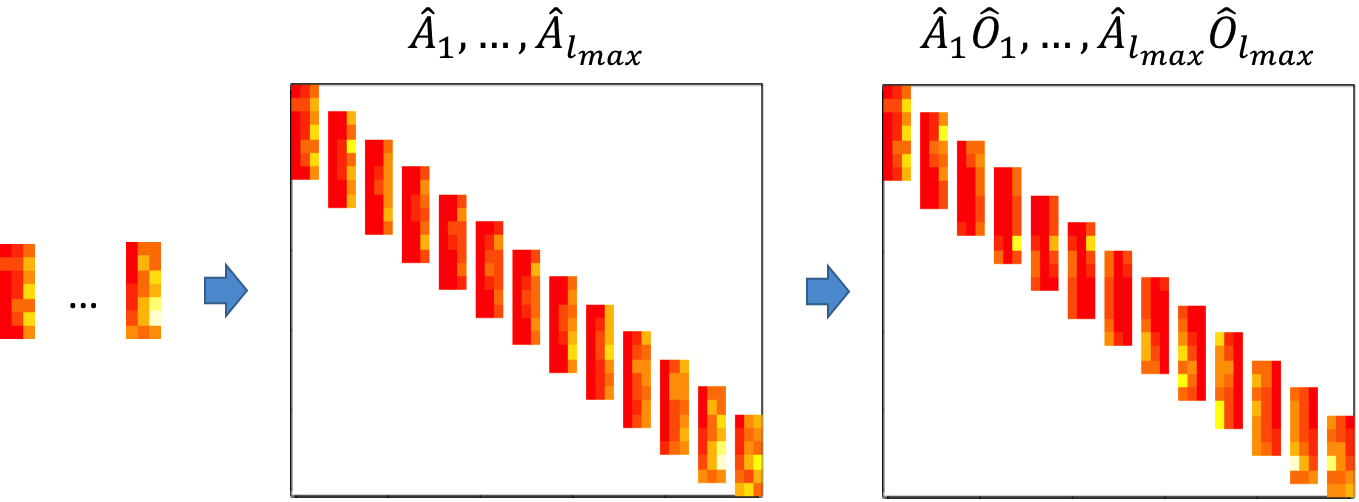}}\\
		\subfigure[Steps 5 and 6. Aggregate $\hat{A}_l$ to $\tilde{A}$ and calculate $\hat{\Sigma}_0 = \tilde{A}\tilde{A}^\top$]{~\quad \includegraphics[height=1.5in]{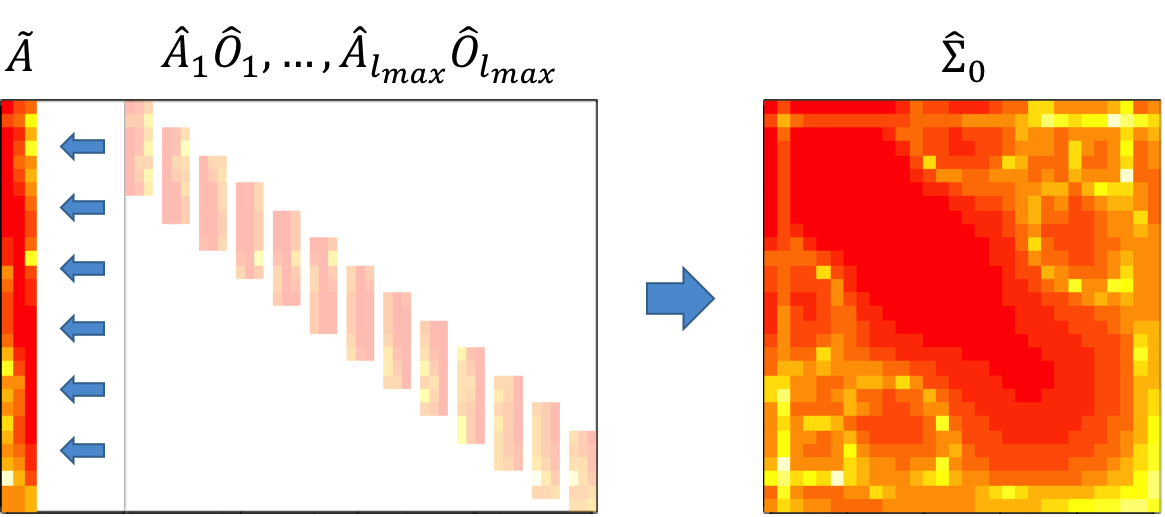}\quad~}\\		
	\end{center}
	\caption{Illustration of the procedure}
\end{figure}

We consider a sequential-aggregation-based algorithm. We first divide $\mathcal{T}$ into a series of overlapping sub-domains, then obtain estimates of $A$ on each sub-domain. Next, we aggregate all estimates on the sub-intervals into a full estimate of $A$. Here, a crucial rotation operation is involved in the aggregation step to ensure that the estimates of $A$ on each sub-domain are aligned.  Finally, we obtain an estimate of $\Sigma_0$ from $\tilde{A}\tilde{A}^\top$, where $\tilde{A}$ is an estimate of $A$ up to a rotation. Then, $G$ is recovered using a standard interpolation technique. The steps are as follows; see Figure \ref{fig:procedure}. For any sub-index set $I\subseteq \{1,\dots,p\}$, we use the notation $T(I) = \{t_i: i \in I\}$ and $(X_k)_{I} = X_k(T(I))$.

\begin{enumerate}
	\item[Step 1] For a chosen band parameter $b$ and an increment parameter $a$ satisfying $1\leq a \leq b-r \leq b\leq d$, we construct the following sub-index set:
	\begin{equation}\label{eq:I_l}
	I_l = \left\{(l-1)a + 1, \ldots, \left\{(l-1)a + b\right\}\wedge p \right\}, \quad l=1,\ldots, l_{\max}.
	\end{equation}
	Here, $l_{\max} = 1+ \lceil (p-b)/a \rceil$ is the total number of sub-index sets. Each $I_l$ except the last one contains $b$ indices, and the last one contains at most $b$ indices.
	
	\item[Step 2] For $l=1,\ldots, l_{\max}$, we search for all samples that have full observations in $I_l$, and denote the set of such samples as $J_l$:
	$$J_l = \left\{1\leq k\leq n: T(I_l) \subseteq T_k\right\}.$$ 
	Then, the sample covariance matrix for the indices in $I_l$ is calculated as
	\begin{equation}\label{eq:hat_Sigma_complete}
	\begin{split}
	\hat{\Sigma}_l\in \mathbb{R}^{|I_l| \times |I_l|}, \quad & \hat{\Sigma}_l = \frac {1}{n_l^\ast}\sum_{k \in J_l}\left((X_{k})_{I_l} - \bar{X}_{I_l}\right)\left((X_{k})_{I_l} - \bar{X}_{I_l}\right)^\top, \\
	& n_l^\ast = |J_l|, \quad \bar{X}_{I_l} = \frac{1}{n_l^\ast}\sum_{k \in J_l}(X_k)_{I_l}.
	\end{split}
	\end{equation}
	
	\item[Step 2']  As an alternative to using only subjects that have full observations in $I_l$, we can use all data available for the pair $(I_l(i), I_l(j))$ when computing $\hat{\Sigma}_{l,[ij]}$. This scheme is preferred to Step 2 when large portions of subjects have missing values; that is, $X_k(T_k)$ are not complete consecutive observations (see Theorem \ref{th:upper_bound_func} and Remark \ref{rm:theorem1-2-comparison}): $\hat{\Sigma}_l\in \mathbb{R}^{|I_l| \times |I_l|}, \bar{X}_{I_l} \in \mathbb{R}^{|I_l|},$
	\begin{equation}\label{eq:hat_Sigma_incomplete}
	\begin{split}
	& \hat{\Sigma}_{l,[ij]} = \frac{\sum_{k: T(I_l(i)), T(I_l(j)) \in T_k} \left((X_k)_{I_l(i)} - \bar{X}_{I_l(i)}\right)\left((X_k)_{I_l(j)} - \bar{X}_{I_l(j)}\right)}{(n^\ast)_{i,j,l}}, \\
	& n^\ast_{i,j,l} = \left|\left\{k: T(I_l(i)), T(I_l(j)) \in T_k\right\}\right|,\\
	& \bar{X}_{I_l(i)} = \frac{\sum_{k: T(I_l(i)), T(I_l(j)) \in T_k} (X_k)_{I_l(i)}}{n^\ast_{i, l}}, \quad n^\ast_{i,l} = \left|\left\{k: T(I_l(i))  \in T_k\right\}\right|.
	\end{split}
	\end{equation}
	
	\item[Step 3] Evaluate the eigenvalue decomposition and the rank-$r$ truncation of $\hat{\Sigma}_l$ as 
	\begin{equation}
	\hat{\Sigma}_l = \hat{U}_l\hat{D}_l\hat{U}_l^\top,\quad \hat{\Sigma}_l^{(r)} = \hat{U}_{l, [:, 1:r]}\hat{D}_{l, [1:r, 1:r]}\hat{U}_{l, [:, 1:r]}^\top.
	\end{equation}
	Then, for $l=1,\ldots, l_{\max}$, we evaluate $\hat{\sigma}_l^2 = (\frac{1}{|I_l|-r}\sum_{i=r+1}^{|I_l|}\hat{D}_{l, [i, i]})\vee 0$ as the sample variance of the noise and 
	\begin{equation}\label{eq:hat-A_l}
	\hat{A}_l = U_{l, [:, 1:r]} \left\{(D_{l, [1:r, 1:r]} - \hat{\sigma}_l^2\cdot I_{r\times r}) \vee 0 \right\}^{1/2} \in \mathbb{R}^{|I_l| \times r}
	\end{equation}
	as the estimate of $A$ on the sub-domain $I_l$. Here, $I_{r\times r}$ is the $r$-by-$r$ identity matrix. By these calculations, we expect that $\hat{A}_l\hat{A}_l^\top \approx \Sigma_{0, l} = (\Sigma_0)_{[I_l, I_l]}$.

	\item[Step 4] We construct a suitable right rotation on $\hat{A}_l$ so that all the pieces can be aligned. Specifically, we first let $\hat{O}_1 = I_{r\times r}$, and then calculate $\hat{O}_{l+1}$ sequentially as
	\begin{equation}\label{eq:hat_O_l}
	\hat{O}_{l+1} = \argmin_{O\in \mathbb{O}_{r}} \left\|(\hat{A}_l)_{[(a+1): b, :]}\hat{O}_l - (\hat{A}_{l+1})_{[1:(b-a), :]} O\right\|_F^2, l = 1,\ldots, l_{\max}-1.
	\end{equation}
	Here, the row indices of $(\hat{A}_l)_{[(a+1): b, :]}$ and $(\hat{A}_{l+1})_{[1:(b-a), :]}$ both correspond to $[la+1, (l-1)a+b]\subseteq\{1,\ldots, p\}$. Note that \eqref{eq:hat_O_l} is actually the orthogonal Procrustes or Wahba problem \citep{wahba1965least}, which can be solved using
	\begin{equation}
	\hat{O}_{l+1} = \tilde{U}\tilde{V}^\top, \text{ where }  \tilde{U}\tilde{\Sigma}\tilde{V}^\top = (\hat{A}_{l+1})_{[1:(b-a), :]}^\top (\hat{A}_l)_{[(a+1): b, :]}\hat{O}_l \text{  is the SVD}.
	\end{equation}
	
	\item[Step 5] In this step, we aggregate all pieces $\hat{A}_l\hat{O}_l$ into one complete factor $\tilde{A} \in \mathbb{R}^{p\times r}$. For convenience of notation, we ``frame" the $|I_l|$-by-$r$ matrix $\hat{A}_l$ to its original $p$-by-$r$ factor scale, $\hat{A}^\ast_l \in \mathbb{R}^{p\times r}$, $\hat{A}^\ast_{l, [I_l, :]} = \hat{A}_l\hat{O}_l$, and $\hat{A}^\ast_{l, [I_l^c, :]} = 0$. For $1\leq i\leq p$ and $1\leq j\leq r$, we calculate
	\begin{equation}\label{eq:tilde-A}
	\tilde{A}_{[i, j]} = \frac{\sum_{l: i\in I_l}\hat{A}_{l, [i, j]}^\ast}{\left|\left\{l: i\in I_l\right\}\right|}.
	\end{equation}	
	\item[Step 6] After the sequential aggregation, we estimate $\Sigma_0$ using 
	\begin{equation}
	\hat{\Sigma}_0 = \tilde{A}\tilde{A}^\top \in \mathbb{R}^{p\times p},
	\end{equation}
	then linear interpolate between grid points to obtain $\hat G$ \citep[Chapter~3.6]{Press:1992:NRC:148286}. Some smoothing instead of linear interpolation might be useful in data applications for smoother results and better visualization.
\end{enumerate}

\noindent{\bf Computation and Tuning Parameters:} In summary, the proposed algorithm is noniterative, and uses only basic matrix calculations, such as matrix multiplications and SVD, which can be implemented efficiently. The algorithm takes input as $a$, $b$, and the rank $r$. According to our simulation studies in Section \ref{sec:numerical}, the performance of the method is not sensitive to the selection of $a$ and $b$. In our numerical implementation, we suggest selecting $b$ to be slightly smaller than bandwidth $d$, and selecting $a$ to be a small increment (in practice $a = 0.1\times d$ usually provides a good enough result). In the following, we describe the random sub-sampling cross-validation method \citep{picard1984cross} used to select the rank $r$. 

We first randomly split $n$ observations $\{X_k(T_k)\}_{k=1}^n$ into training and testing groups of sizes $n_1 \approx \frac{(K-1)n}{K}$ and $n_2 \approx \frac{n}{K}$, respectively, $T$ times. For the $t$th split, let $J_{\rm train}^{(t)}$ and $J_{\rm test}^{(t)}$ be the index sets for the training and testing groups, respectively. For each $r \in \{1,\ldots,b-a\}$, we apply the proposed procedure to the training dataset $\{X_k(T_k)\}_{k\in J_{\rm train}^{(t)}}$, and denote the outcome as $\hat{\Sigma}^{(t)}(r)$. Then, we calculate the sample covariance matrix $\hat{\Sigma}_{\rm test}^{(t)}\in \mathbb{R}^{p\times p}$ based on the samples from the testing group,
\begin{equation*}
\begin{split}
& (\hat{\Sigma}_{\rm test}^{(t)})_{[i, j]} = \\
& \left\{\begin{array}{ll}
\sum_{\substack{k \in J_{\rm test}^{(t)}\\T_k\ni T(i), T(j)}} (X_{ik} - \bar{X}_i) (X_{jk}-\bar{X}_j)\bigg/  \sum_{\substack{k \in J_{\rm test}^{(t)}\\ T_k\ni T(i), T(j)}}  1, & \text{if } \sum_{\substack{k \in J_{\rm test}^{(t)}\\  T_k\ni T(i), T(j)}}  1 \geq n_0,\\
{\rm NA}, &  {\rm otherwise},
\end{array}
\right. 
\end{split}
\end{equation*}
where $n_0$ is the lower threshold when evaluating the testing sample covariance matrix. Then, we evaluate the prediction error as
$$E(r) = \sum_{t=1}^T \sum_{(\hat{\Sigma}_{\rm test}^{(t)})_{[i, j]} \neq {\rm NA}} \left((\hat{\Sigma}^{(t)}(r))_{[i, j]} - (\hat{\Sigma}_{\rm test})_{[i, j]}\right)^2.$$
Here, to improve accuracy, we only evaluate the prediction errors on those $(i, j)$ pairs where $(\hat{\Sigma}_{\rm test})_{[i, j]}$ is evaluated based on at least $n_0$ samples. Finally, we choose $\hat{r}  = \argmin_{1\leq r \leq b-a} E(r)$, and apply the proposed procedure with $\hat{r}$ to obtain the final estimator $\hat{\Sigma}_0$. In our simulations, we use $K= 5$, $T= 10$, and $n_0 = 4$; other cross-validation methods are expected to yield similar results.

In practice, we propose using the cross-validation method, because this usually prevents under-selection. We observed a slight over-selection of $r$ in our simulations, but this is not a problem in a covariance estimation because the components (eigenvalues) beyond $r$ are all assumed to be very small. In Section \ref{sec:numerical}, we examine the numerical performance of the proposed procedure based on cross-validation and the effect of the tuning parameters.

\section{Theoretical Analysis}\label{sec:theory}

Before presenting the main theoretical results, we first introduce the following assumptions.

\begin{Assumption}\label{as:lowrank} 
	There is a positive integer $r$ such that the eigenvalues of $G$ satisfy $ \lambda_1(G) \geq \dots \geq \lambda_r(G) > \lambda_{r+1}(G) \geq \cdots \geq 0$. Let $G^{(r)}$ be the best rank-$r$ approximation for $G$ and $G^{-(r)} = G - G^{(r)}$. We also assume $\|G\|_{HS} < \infty$, $\|G^{(-r)}\|_{HS} \leq \frac{C}{\sqrt{n^\ast}}$, where $n^\ast$ is the effective sample size, defined in Theorem \ref{th:upper_bound_func}. 
\end{Assumption}

The rank $r$ is allowed to increase slowly as $n$ and $p$ grow. The (approximate) reduced-rank covariance structure is explored by \cite{james2000principal} and \cite{peng2009geometric} for sparse functional data, where only a few irregularly (randomly) spaced observations are available on each subject. They view the rank restriction as a form of regularization to avoid over-parametrization. The same reasoning applies to our scenario, because only a fraction of the trajectories are observed for each subject.

\begin{Assumption}\label{as:sub-matrix}
	For any contiguous subdomain $\mathcal{\tilde T}\subseteq \mathcal{T}$, we define $G^{(r)}_{[\tilde{\mathcal{T}}, \tilde{\mathcal{T}}]} = G^{(r)}(s,t)_{s\in  \mathcal{\tilde T}, t\in  \mathcal{\tilde T}}$. There exists a constant $0< \kappa < \delta$, such that \\$\gamma = \max_{\frac{\mathcal{L}(\mathcal{\tilde T})}{\mathcal{L}(\mathcal{T})}\geq \kappa}\{\tr(G)\frac{\mathcal{L}(\mathcal{\tilde T})}{\mathcal{L}(\mathcal{T})}/\lambda_r\left(G^{(r)}_{[\tilde{\mathcal{T}}, \tilde{\mathcal{T}}]}\right)\}$ satisfies $\gamma = o((n^{*})^{1/2})$.
\end{Assumption}
Intuitively speaking, this assumption imposes a lower bound of $C/\gamma$ on the $r$th eigenvalues of $G_{[\tilde{\cT}, \tilde{\cT}]}$. It essentially ensures that $G$ restricted on different contiguous subdomain $[\tilde{\cT}, \tilde{\cT}]$ is nonsingular, so that being able to estimate $G$ using only segments of the functional observations is possible. As a counter-example, if $G$ has two ``spikes" in the sense that only $G_{[0:0.2, 0:0.2]}$ and $G_{[0.8:1, 0.8:1]}$ have significant amplitudes, while $G_{[0.2:0.8, 0.2:0.8]}$ is zero, then the estimation of the cross-covariance parts $G_{[0:0.2, 0.8:1]}$ and $G_{[0.8:1, 0:0.2]}$ is impossible when one can only observe functional segments of length no more than $0.6$.  In addition, $\gamma$ is allowed to increase moderately as $n$ and $p$ grow. Note that $\gamma \geq r$, and in the scenarios in which $\gamma/r$ is big, the method using complete observations only (step 2) is better than step 2'.

\begin{Assumption} \label{as:sub-gaussian}
	Assume $X$ satisfies the moment condition $\sup_t\mathbb{E}|X(t)|^4 \leq C$.
\end{Assumption}

\begin{Assumption}\label{as:lipschitz}
	There exists $L>0$, such that $|G(s,t)-G(s',t')|\le L\max(|s-s'|,|t-t'|)$, for all $s,s',t,t'\in \mathcal{T}.$
\end{Assumption}
Because we use sample covariance approach and interpolate between observed grid points, the Lipschitz condition is almost necessary. It is easy to satisfy because we work with a finite domain $\mathcal{T}$, and it is weaker than the second differentiable conditions usually used in smoothing methods. 

We can now state the main results of this study. 

\begin{Theorem}\label{th:upper_bound_func}
	Suppose Assumptions 1--4 hold. We take $b = \beta p, a = \alpha p$ for some constants $0<\alpha < \beta \leq \delta < 1$. Assume $\beta-\alpha \geq \kappa\geq 2r/p$ ($\kappa$ and $r$ were defined in the assumptions), $n \geq C p$, and $p \geq C\gamma$. Then, the proposed procedure yields
	\begin{equation}\label{eq:hat_G-G-complete}
	\mathbb{E}\|\hat{G} - G\|_{HS} =  O\left(\sqrt{\gamma^2/n^\ast}  + p^{-1}\right).
	\end{equation}
	Here,  $n^\ast = \min_{l}n_l^\ast$ and $n_l^\ast$ is defined in \eqref{eq:hat_Sigma_complete}. If we use complete samples to calculate $\hat{\Sigma}_l$ using \eqref{eq:hat_Sigma_complete} of Step 2; $n^\ast = \min_{i, j, l}n^{\ast}_{i, j, l}$ and $n^\ast_{i, j, l}$ is defined in \eqref{eq:hat_Sigma_incomplete} if we use both complete and incomplete samples to calculate $\hat{\Sigma}_l$ using \eqref{eq:hat_Sigma_incomplete} of Step 2'.
	
\end{Theorem}

\begin{Remark}\rm
	The first error term in \eqref{eq:hat_G-G-complete} is due to estimating errors of the discretized covariance $\Sigma_0$. The second error term $p^{-1}$ is from the linear interpolation of the discretized $\Sigma_0$. 
\end{Remark}

\begin{Remark}\rm 
	Theorem \ref{th:upper_bound_func} provides theoretical guarantees for the proposed procedure under general mixed longitudinal designs (conditional on $T_k$), where the effective sample size, $n^\ast$, is driven by the minimum number of samples that cover each sub-interval $I_l$. In a balanced design, where $T_k \subseteq T(\{w_k, \dots, w_k+d-1\})$, with $w_k$ evenly chosen from $\{1, \ldots, p-d+1\}$, for $k=1,\ldots, n$, the boundary sub-intervals $I_1$ and $I_{l_{\max}}$ have less effective sample sizes than those of middle ones, which yield a higher estimation error for the boundary part of $G$. To overcome this bottleneck, we recommend a boundary-enriched design: beyond the balanced design, as mentioned above, we include $n_a = cn$ additional ones with $T_k=T(\{1,\dots, d\})$ or $T(\{p-d+1,\dots,p\})$ for a small constant $0 \le c \le 1$. 
	Alternatively, one can apply an extended-domain design: for each $k=1,\dots, n$, $T_k = T(\{w_k,\dots, w_k+d-1\}) \cup T(\{1,\dots, p\})$, with $w_k$ uniformly chosen from $\{(2-d), \ldots, p\}$. Under both the boundary-enriched and the extended-domain designs, the result of Theorem \ref{th:upper_bound_func} yields \eqref{eq:hat_G-G-complete}.
\end{Remark}

\begin{Remark}[Proof sketch of Theorem \ref{th:upper_bound_func}]\rm 
	After introducing some notation, we develop error bounds for $\hat{A}_l$ (the outcome of Step 3), $\hat{O}_l$ (the outcome of Step 4), $\tilde{A}$ (the outcome of Step 5), $\hat{\Sigma}_0$, and the final estimator $\hat{G}$ (the outcome of Step 6). In particular, Step 4 of the proposed procedure involves solving the orthogonal Procrustes problem (or Wahba problem) \eqref{eq:hat_O_l}. To derive the error bound of $\hat{O}_l$ from the error bound of $\hat{A}_l$, we introduce  \ref{lm:rotations-Wahba}, which provides a theoretical guarantee for the solution of \eqref{eq:hat_O_l}. In addition, Lemma \ref{lm:rotations-Wahba} is stronger than previous results (cf., \cite[Lemma 16]{bishop2014deterministic}), which may be of independent interest. 
	\begin{Lemma}[Perturbation bound for Wahba problem]\label{lm:rotations-Wahba}
		Suppose $A_1, A_2, A\in \mathbb{R}^{m \times r}$, $O_1, O_2 \in \mathbb{O}_{r}$, $\|A_1 - AO_1\|_F \leq a_1, \|A_2 - AO_2\|_F\leq a_2$, and $\sigma_r(A)\geq \lambda$. Suppose $\hat{O}$ is the solution to Wahba problem,
		\begin{equation*}
		\begin{split}
		& \quad \hat{O} = \argmin_{O\in \mathbb{O}_r}\left\|A_2 O - A_1\right\|_F,\\ \text{or equivalently}, &  \quad \hat{O} = UV^\top \text{ if } A_2^\top A_1 = U\Sigma V^\top \text{ is the SVD}.
		\end{split}
		\end{equation*}
		Then, $\hat{O}$ satisfies
		\begin{equation}
		\left\|\hat{O} - O_2^\top O_1 \right\|_F \leq \frac{2(a_1+a_2)}{\lambda}. 
		\end{equation}
	\end{Lemma}
\end{Remark}

Proposition \ref{pr:complete} provides a sharper convergence rate when Step 2 is applied (with only complete pieces) and the random scores are sub-Gaussian distributed.

\begin{Proposition}\label{pr:complete}
	Suppose $Z(t) = \mu(t) + \sum_{k \geq 1}\xi_k\phi_k(t)$ is the Karhunen--Lo\`eve decomposition, where $\{\phi_k(t)\}_{k\geq 1}$ is the fixed eigenfunction and $\{\xi_k\}_{\geq 1}$ are random scores. In addition to the assumptions of Theorem 1, we further assume the normalized leading $r$ scores, $\tilde{\xi} = \{\xi_k/\lambda_k^{1/2}(G)\}_{k=1}^r$, are sub-Gaussian distributed, such that $\mathbb{E}\exp(t\tilde{\xi}^\top u) \leq \exp(C\|u\|_2^2)$, for any $u\in \mathbb{R}^r$, the tail part $Z^{(-r)}(t) = \sum_{k \geq r+1} \xi_k\phi_k(t)$ satisfies $\sup_t\mathbb{E}|Z^{(-r)}(t)|^4 \leq Cr/(n^\ast \gamma)$, and the noise satisfies $(\mathbb{E}|\epsilon|^4)^{1/2} \leq Cr/\gamma$. Then, the proposed procedure with Step 2 yields the following rate of convergence:
	\begin{equation}\label{ineq:sharper-rate}
	\mathbb{E}\|\hat{G} - G\|_{HS} = O\left(\sqrt{r\gamma/n^\ast} + p^{-1}\right).
	\end{equation}
	Here,  $n^\ast = \min_{l}n_l^\ast$ and $n_l^\ast$ is defined in \eqref{eq:hat_Sigma_complete}.
\end{Proposition}

\begin{Remark}
	\label{rm:theorem1-2-comparison}\rm 
	We briefly compare the convergence rates of step 2 and step 2'. First, $n^\ast$ when using the complete sample is no greater than that when using both complete and incomplete subjects.  On the other hand, the factor $\gamma^2$ in \eqref{eq:hat_G-G-complete} is greater than $r\gamma$ in the counterpart of \eqref{ineq:sharper-rate}. This is because $\hat{A}_l$ calculated using the standard sample covariance matrix, as in Step 2, possesses a sharper convergence rate than that calculated using the extended sample covariance matrix, as in Step 2', as demonstrated by Lemma \ref{lm:factorization-lemma}. Therefore, there is a trade-off between using Steps 2 or 2'. In general, we recommend using Step 2' when most subjects have noncontiguous observations (missing values); otherwise Step 2 is preferred. 
\end{Remark}

\section{Numerical Experiments}\label{sec:numerical}

\noindent{\bf Simulations:} In this section, we investigate the numerical performance for the proposed procedure using a series of simulation studies. For each setting, we generate $X_{ij} = \sum_{k=1}^{K}\xi_{ik}\phi_{k}(t_{ij}) + \epsilon_{ij}$, where $i=1,\ldots, n$, $j=1,\ldots, p$, and $t_{ij}$ are equally spaced $p$ values on $[0,1]$. We observe a contiguous $(\delta = d/p)$ portion of the trajectory for each subject. All simulation results are based on 100 repetitions.

\begin{figure}
	\centerline{
		\includegraphics[width = 1.9in, height = 2in]{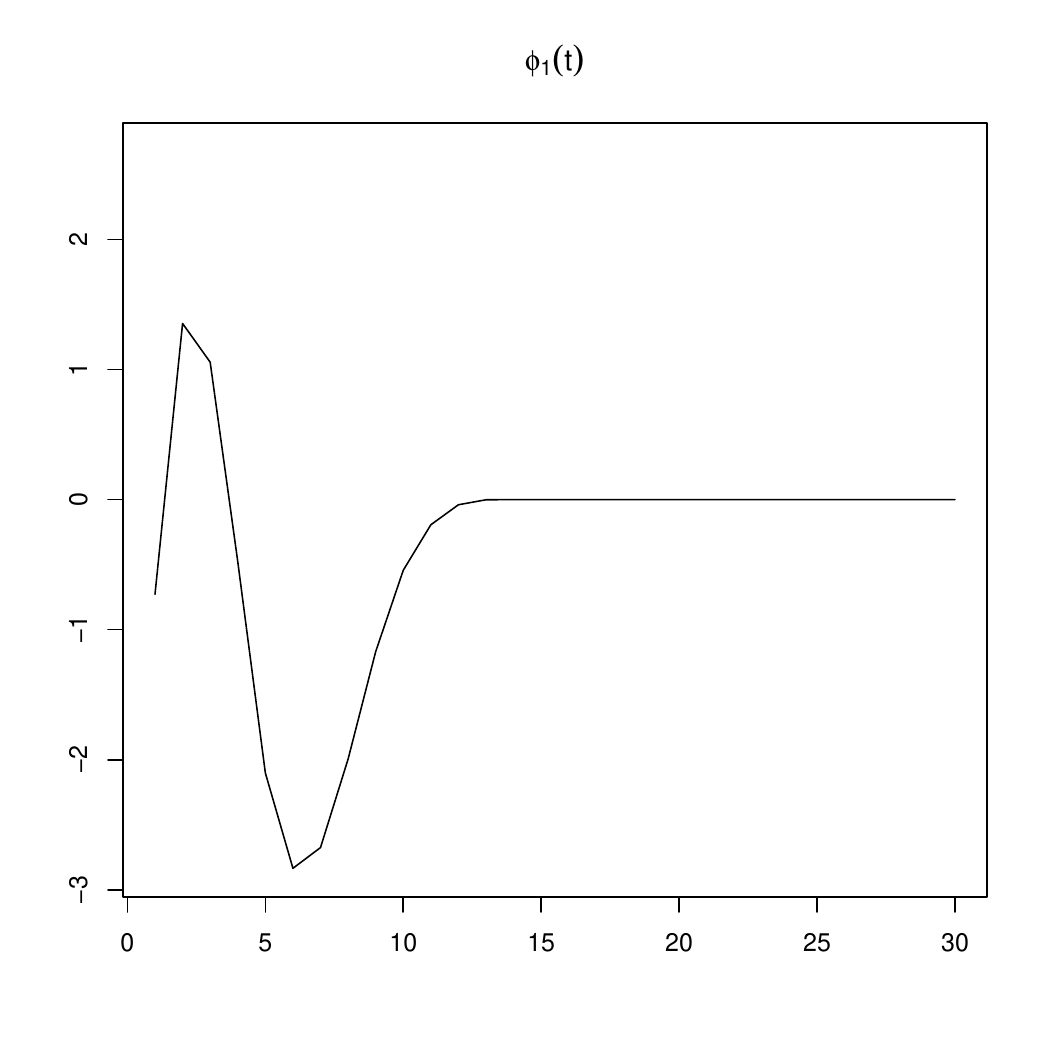}\includegraphics[width = 1.9in, height = 2in]{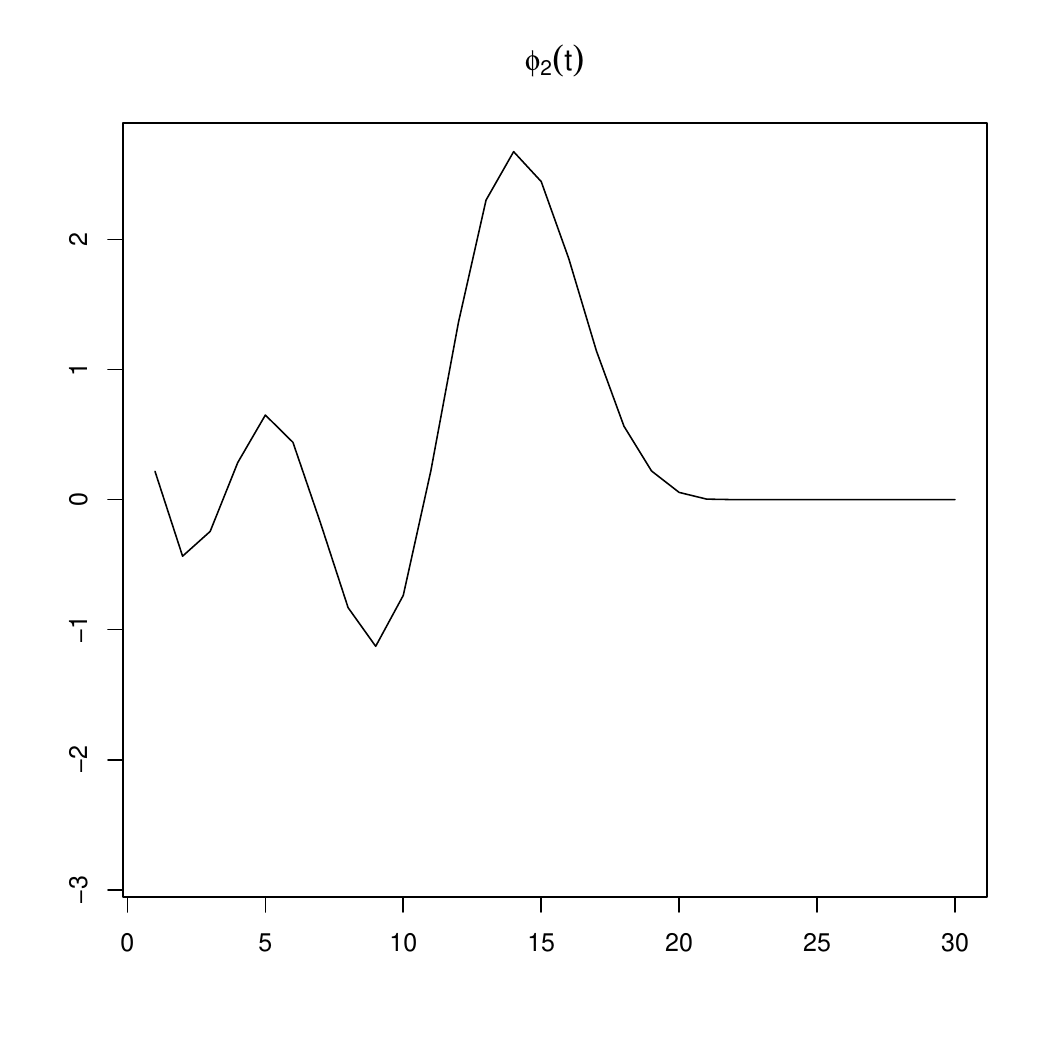}	\includegraphics[width = 1.9in, height = 2in]{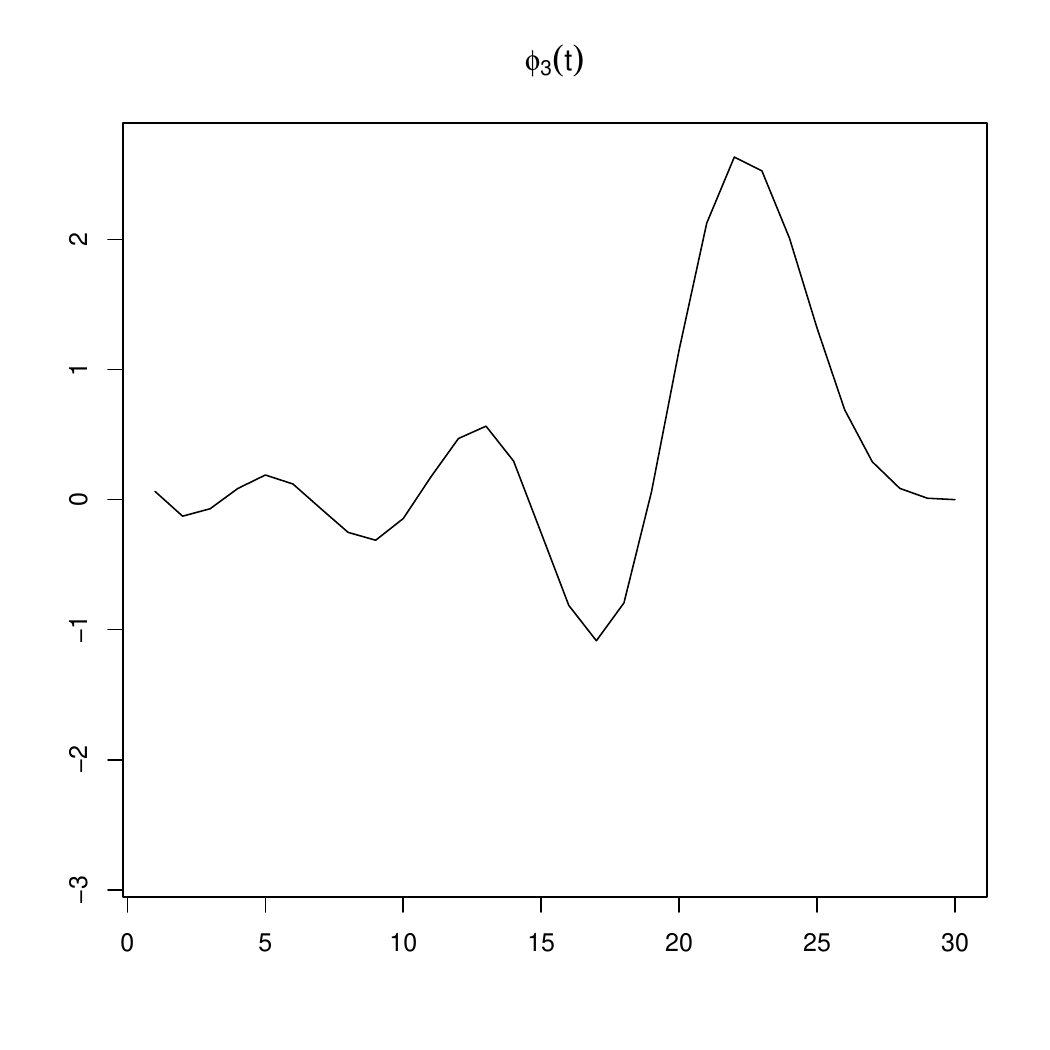}}
	\caption{The first three eigenfunctions used in the simulations to generate the data.}
	\label{fig:eigenfunctions} 
\end{figure}

The first simulation setting is designed to assess the basic performance of the proposed method, and to explore the choices of the tuning parameters. In particular, we set $p = 30$, the true rank $K= 3$, and the eigenfunctions $\{\phi_k(t)\}$ as linear combinations of $M=10$ cubic B-splines with equally spaced knots, as shown in \Cref{fig:eigenfunctions}. The random scores $\{\xi_{ik}\}$ are i.i.d normal with variances $(\lambda_1, \lambda_2,\lambda_3) = (4^2, 3^2, 2^2)$. The errors $\epsilon_{ij}$ are i.i.d normal with variance one.  We let the length of the observation band $d=10$, so that each observation band covers one-third ($d/p$) of the total domain. We further let each contiguous subset of length $d$ be observable by $n_{rep} = \{10, 20, 50\}$ subjects, which means the total sample size $n = n_{rep}\times(p-d+1) = 210, 420, 1050$. We apply the proposed method in Section 2, with the rank $r$ selected by cross-validation, as described in Section \ref{sec:method}, and report the relative estimation errors for different choices of tuning parameters $b$ and $a$ in \Cref{tab:simu1}. Here, the relative estimation error in all simulation settings is defined as $\|\hat{G} - G\|_{HS}/\|G\|_{HS}$. We can see that the estimation error decreases as the sample size increases, and the performance is not sensitive to the values of ($b,a$), as long as $b$ is slightly smaller than the bandwidth $d$ and $a$ is small. The cross-validation of the proposed method tends to slightly over-select $r$, but over-selection does not affect the RMSE of the covariance estimation significantly in our simulation settings. In the following simulations, we always use the bandwidth $b= \ceil{0.7\cdot d}$ and the incremental parameter $a = \ceil{0.1\cdot d}$

\begin{table}[ht]
	\caption{Results for simulation 1: the average relative error over 100 simulations are shown, with the standard error given in parentheses. Here, $a$ and $b$ are different choices of tuning parameters, and the results are stable. }
	\vskip 5pt
	\label{tab:simu1}
	\centering
	\vskip 2pt
	\begin{tabular}{r|lll}
		\hline
		\hline
		& $n_{rep}=10$ & $n_{rep}= 20$ & $n_{rep} = 50$  \\ 
		\hline
		$b=7,a=1$ & 0.324 (0.17) & 0.224 (0.14) & 0.123 (0.06) \\ 
		$b=7,a=2$  & 0.325 (0.16) & 0.221 (0.13) & 0.132 (0.1) \\ 
		$b=8,a=1$ & 0.314 (0.17) & 0.23 (0.14) & 0.13 (0.09) \\ 
		$b=8,a=2$ & 0.364 (0.17) & 0.292 (0.19) & 0.126 (0.08) \\ 
		$b=9,a=1$ & 0.326 (0.16) & 0.227 (0.15) & 0.119 (0.07) \\ 
		$b=9,a =2$ & 0.347 (0.15) & 0.214 (0.11) & 0.145 (0.11) \\ 
		\hline
	\end{tabular}
\end{table}

The second simulation setting further explores the performance under different settings. In particular, let $p=30$ and the fraction of observable domain $\delta = \{1/5, 1/3, 1/2\}$. In addition to the previous setting with $K=3$, we also consider $K=10$, the score variances $(\lambda_1,\ldots, \lambda_{10}) = (4^2, 3^2, 2^2, 2^{-4}, \ldots, 2^{-10})$, 
$\phi_1, \phi_2, \phi_3$ are the same as in the previous settings, and $\phi_k(t) = \sqrt{2}\sin(k\pi t)$, for $k = 4, \dots, 10$ (all 10 functions are orthonormalized). 
Similarly to the first simulation setting, we implement the proposed procedure with $r$ selected using cross-validation, and let $b= \ceil{0.7\cdot d}$ and $a = \ceil{0.1\cdot d}$; see \Cref{tab:simu2}. We can see that the proposed procedure still performs well when there are moderate deviations to the reduced-rank structure. The estimation error decreases as the observed partial trajectory covers a larger fraction of the entire trajectory. 
Note that the selected rank $r$ for the cases $K=10$ increases as the sample size increases, with an average value $r = 4.25$ for $d/p = 1/3$ and $n_{rep} = 50$.

\begin{table}[ht]
	\caption{Results for simulation 2: the average relative error over 100 simulations are shown, with the standard error given in parentheses. Here, $K$ is the total number of eigenfunctions used to generate the covariance, and $\delta$ denotes the fraction of domains observed.}
	\vskip 5pt
	\label{tab:simu2}
	\centering
	\vskip 2pt
	\scalebox{0.81}{
		\begin{tabular}{r|lll|lll}
			\hline
			\hline
			&  \multicolumn{3}{c|}{$K=3$}&\multicolumn{3}{c}{$K=10$}\\
			&  $n_{rep} = 10$ & $n_{rep}= 20$ & $n_{rep} = 50$ & $n_{rep} = 10$ & $n_{rep}= 20$ & $n_{rep} = 50$ \\ 
			\hline
			$\delta = 1/5$ &  0.43 (0.17) & 0.397 (0.2) & 0.294 (0.21) & 0.461 (0.16) & 0.403 (0.18) & 0.304 (0.19)  \\ 
			\hline
			$\delta=1/3$  &0.341 (0.17) & 0.237 (0.16) & 0.135 (0.1) & 0.322 (0.16) & 0.248 (0.14) & 0.143 (0.06)  \\ 
			\hline
			$\delta=1/2$  &0.243 (0.11) & 0.17 (0.07) & 0.113 (0.05) & 0.248 (0.1) & 0.165 (0.05) & 0.114 (0.04) \\ 
			\hline
		\end{tabular}
	}
\end{table}

The third simulation explores the performance when there are further missing values within the observable fraction of the domain. 
The setting is the same as that in the first simulation, except that the data have a 5\%, 10\%, or 15\% missing rate. As in the previous two simulations, we implement the proposed procedure with $r$ selected by cross-validation, and let $b= \ceil{0.7\cdot d}$, and $a = \ceil{0.1\cdot d}$; see \Cref{tab:simu3}. We can see that the proposed procedure performs reasonably well when there is a moderate number of missing values, and the performance improves when the sample size becomes large. 

\begin{table}[ht]
	\centering
	\caption{Results for simulation 3: the average relative error over 100 simulations are shown, with the standard error given in parentheses. Here, ``missing" is the percentage of missing values within the observed domain. }
	\label{tab:simu3}
	\vskip 5pt
	\begin{tabular}{r|llll}
		\hline
		\hline
		missing & $n_{rep}=10$ & $n_{rep}=20$ & $n_{rep} = 50$ & $n_{rep} = 100$ \\ 
		\hline
		5\%& 0.36 (0.14)  & 0.24 (0.12)  & 0.16 (0.07)  & 0.12 (0.06)  \\ 
		10\% & 0.39 (0.16)  & 0.29 (0.13)  & 0.19 (0.08)  & 0.13 (0.05)  \\ 
		15\% & 0.42 (0.13)  & 0.32 (0.13)  & 0.22 (0.1)  & 0.16 (0.06)  \\ 
		\hline
		\hline
	\end{tabular}
\end{table}

The fourth simulation compares the performance of the proposed method with the matrix completion method proposed in \cite{descary2017recovering}. The data-generating procedure is the same as those of the previous simulations. The matrix completion method is implemented using the Matlab code downloaded from the authors' website. The method requires an input of rank $r$, and they propose using a scree-plot to manually determine the rank (looking for an ``elbow" in the plot). Because this approach is not feasible in simulation settings, we use the true rank $r$ for both methods. 
The results are reported in \Cref{tab:simu4}. The relative performance depends on the fraction of the domain observed. For $\delta = 1/2$, both methods work fine, and the matrix completion method is slightly better for a small sample size ($n_{rep}= 10$). For $\delta = 1/3$, both methods work fine, and the proposed method is slightly better for larger sample sizes. For $\delta = 1/5$, neither of the methods work well for a small sample size ($n_{rep} = 10$), although the error for the matrix completion method is not as large as that of the proposed method. When $n$ increases, the error of the proposed method decreases to a reasonably small level; the matrix completion method is less satisfactory in this case.

\begin{table}[ht]
	\centering
	\caption{Results for simulation 4: the average relative error over 200 simulations are shown, with the standard error given in parentheses. Here, ``MatComp" is the matrix completion method proposed in \cite{descary2017recovering}, and $\delta$ denotes the fraction of domains observed.}
	\vskip 5pt
	\label{tab:simu4}
	\begin{tabular}{r|rllll}
		\hline
		\hline
		& & $n_{rep}=10$ & $n_{rep}=20$ & $n_{rep} = 50$ & $n_{rep} = 100$ \\ 
		\hline
		$\delta = 1/5$ & proposed & 0.37 (0.11)  & 0.27 (0.09)  & 0.17 (0.06)  & 0.12 (0.04)   \\ 
		& MatComp & 0.32 (0.06)  & 0.27 (0.04)  & 0.23 (0.02)  & 0.22 (0.02)  \\ 
		\hline
		$\delta = 1/3$& proposed& 0.26 (0.10)  & 0.2 (0.06)  & 0.12 (0.04)  & 0.08 (0.03)  \\ 
		& MatComp & 0.29 (0.08)  & 0.2 (0.05)  & 0.14 (0.03)  & 0.11 (0.02) \\ 
		\hline
		$\delta = 1/2$& proposed & 0.26 (0.11)  & 0.18 (0.07)  & 0.12 (0.05)  & 0.08 (0.03)  \\ 
		& MatComp & 0.24 (0.08)  & 0.17 (0.05)  & 0.11 (0.03)  & 0.08 (0.02) \\ 
		\hline
		\hline
	\end{tabular}
\end{table}

\begin{figure}
	\centerline{
		\includegraphics[width = 3in, height = 3in]{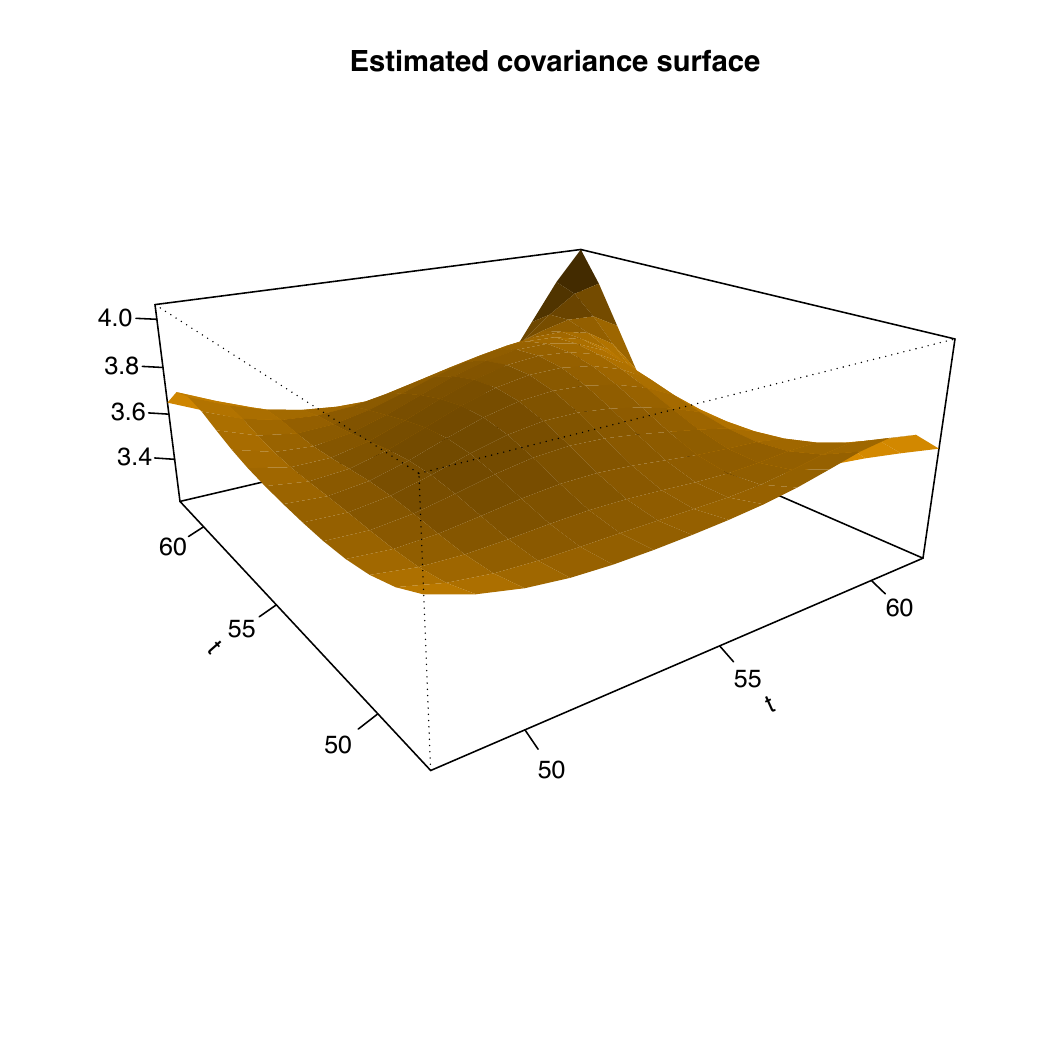}\includegraphics[width = 3.2in, height = 3in]{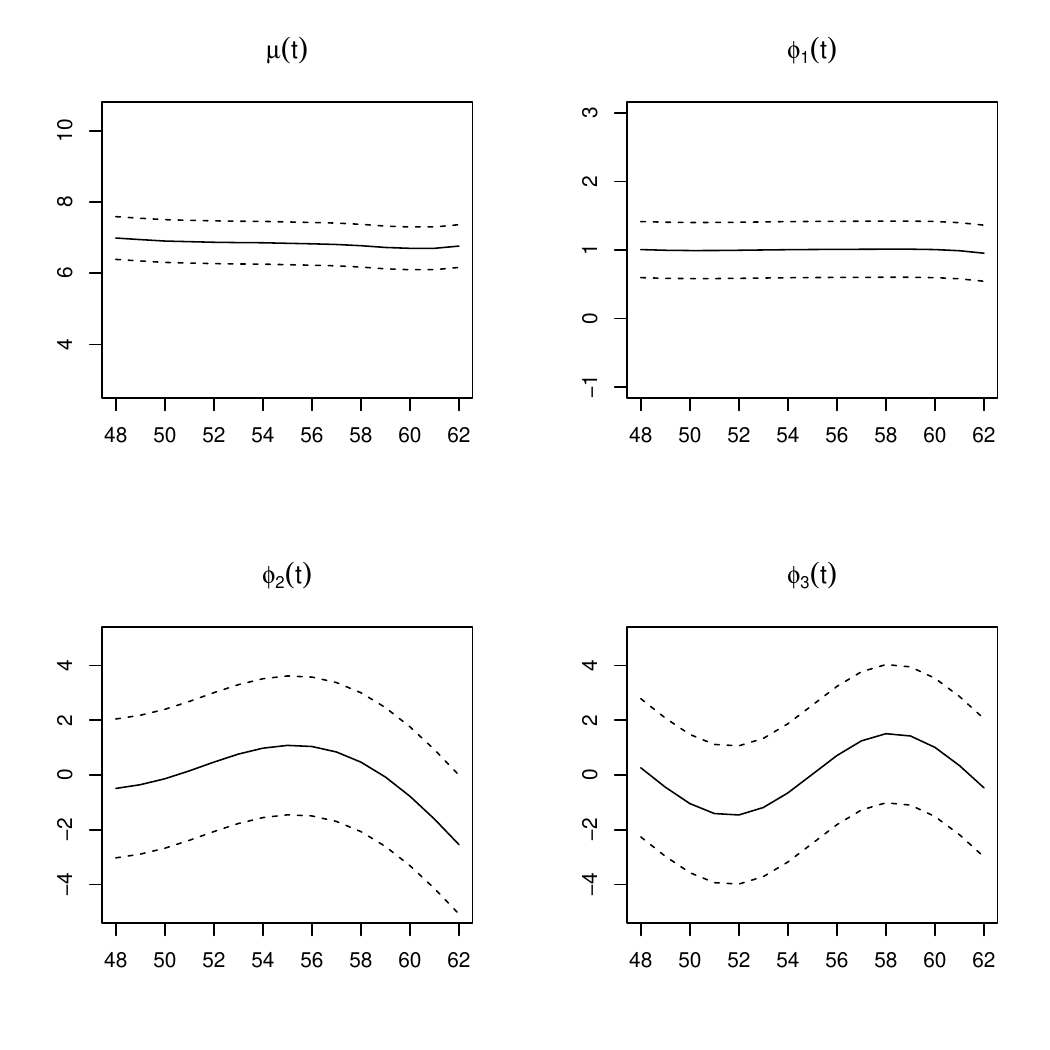}}
	\caption{Left: The estimated covariance surface of the working memory data for women aged between 48 and 62. Right: The estimated mean function and estimated eigenfunctions corresponding to the largest three modes of variation, where the dashed lines are 95\% bootstrap simultaneous confidence bands. }
	\label{Fig:data} 
\end{figure}

\noindent{\bf Application to a study on the working memory of midlife women:} 
We downloaded the data from the SWAN database (link: \url{http://www.icpsr.umich.edu/icpsrweb/ICPSR/series/00253}). The study examines the physical, biological, psychological, and social health of women during their middle years. In this section, we focus on the measurement of working memory, that is, the ability to manipulate information held in memory. In this study, working memory was assessed using digit span backwards (DSB) \citep{WMmanual}: participants repeat strings of single-digit numbers backwards, with two trials at each string length, increasing from two to seven, stop after errors in both trials at a string length; score as the number of correct trials (range, 0--12).  The testing was first administered at the fourth follow-up to 2709 women, and then repeated in the sixth and subsequent visits. The data up to the tenth visit are publicly available. We exclude those subjects who dropped out before the tenth follow-up visit, leaving a sample size of $n=2016$. Following previous literature, we did not use the first measurement in order to alleviate the practice effect on the testing results \citep{karlamangla2017evidence}. 
Instead, we focused on the age range $\mathcal{T} = [48, 62]$. Each subject has up to five years of consecutive data, and the average number of follow-ups is 3.3. We applied the proposed method described in Section 2 to estimate the covariance function, using a rank $r= 3$ selected by cross-validation, a band parameter $b = 4$, and an increment parameter $a = 1$. The estimated covariance surface is shown in the left panel of Figure \ref{Fig:data}. We can see that the variance is bigger at the middle part around age 55.

The nonparametric covariance estimation serves as a stepping stone for further functional data analysis. 
In the following, we perform a functional principal component analysis for the working memory trajectories, and examine how the shapes of the trajectories depend on education (less than high school, high school, some college/technical school, college graduate, postgraduate), controlling for race (Black, Chinese, Japanese, Caucasian/White, Non-Hispanic, Hispanic) and difficulty paying for basics (no hardship, somewhat hard, very hard). These are just for illustration of the functional data methods; a thorough analysis for this complex data set is beyond the scope of this study. 

Given the estimated covariance, we conducted a functional principal component analysis based on the Karhunen--Lo\`eve expansion $Z(t) = \mu(t) + \sum_j \xi_j\phi_j(t)$. Here, $\{\phi_j(t), j\geq 1\}$ is an orthonormal basis that consists of eigenfunctions of $G$, and $\{\xi_j = \int (Z(t)-\mu(t))\phi_j(t)dt: j\geq 1\}$ are (random) scores. Intuitively, the first $K$ terms expansion, $\mu(t) + \sum_{j=1}^{K} \xi_j\phi_j(t)$, forms a $K$-dimensional representation of $Z(t)$ with the smallest unexplained variance.  
The smoothed mean function and the first three estimated eigenfunctions $\{\phi_j(t), j = 1, 2, 3\}$ are visualized in the right panel of Figure \ref{Fig:data}. We also constructed 95\% confidence bands for these quantities using the nonparametric bootstrap method, as outlined in \cite{hall2006properties}. The best linear prediction methods, as used in \cite{yao2005functional}, were applied to obtain estimates of $\xi_{j}$.

The mean function shows that the working memory function for a middle-age woman is, on average, decreases as she gets older. With longitudinal declines, on average, there are individual differences in working memory aging and possible improvements in performance over multiple years. The first eigenfunction $\phi_1(t)$ is close to a horizontal line. Therefore, $\phi_1(t)$ can be interpreted as a size component: subjects with a positive score in the direction of this eigenfunction have better working memory function than that of an average woman for all ages between 48 and 62. The regression analysis show that this component is significantly and positively correlated with education level, which means that people with higher education tend to have higher working memory scores over the entire period. The other two covariates, financial status and race, are also statistically significant.
The second eigenfunction $\phi_2(t)$ has a reversed U-shape with a maximum at around age = 55. This can be interpreted as a changing pattern before and after age 55, which possibly relates to the menopausal transition, resilience, and compensatory mechanisms \citep{fuh2006longitudinal,greendale2009effects,hahn2015everyday}. Subjects with a positive score in the direction of this eigenfunction have an increase in working memory before age 55, and a fast decline after age 55. The regression analysis show that education is a significant factor, with the postgraduate education group having a more prominent reversed U-shape pattern. The other two covariates are not statistically significant. 
The third component $\phi_3(t)$ crosses the zero line around age 55, representing a complementary effect to the second component. 

This functional data analysis perspective differs from that of traditional linear mixed effect models, because the modes of variation for individual chronological aging trajectories are extracted nonparametrically from the data (FPC components), and one can examine how the shape of the trajectories interact with other covariates. In comparison, traditional linear mixed effect models \citep{karlamangla2017evidence} often control these covariates as fixed main effects.

\section{Conclusions}\label{sec:discussion}

We have focused on data observed on a regular equally spaced grid. The proposed sequential aggregating method can be readily extended to the setting in which the observational times are irregular and random. However, adjustments need to be made to step 2. In particular, the sample covariance estimate for $\Sigma_{l}$ in step 2 is not applicable if the data are irregularly observed. In this case, one can first adopt a bivariate local linear smoothing method \citep{yao2005functional} to estimate the covariance on the observable part (the diagonal banded area), say $\tilde G(s,t)$, for $|s-t|<\delta$. Then, for each piece $l$, take the corresponding sub-piece from $\tilde G(s,t)$, evaluate that on a predefined regular grid $I_l$, and use that as $\hat\Sigma_{l}$. All other steps remain the same.

\section*{Acknowledgments}

The authors thank the editor and two anonymous referees for their helpful comments.

\bibliographystyle{apa}
\bibliography{reference}

\appendix
\newpage

\setcounter{page}{1}
\setcounter{section}{0}

\begin{center}
	{\Large Supplement to ``Nonparametric covariance estimation for mixed longitudinal studies, with applications in midlife women's health"}

	\bigskip\medskip
	{Anru R. Zhang ~ and ~ Kehui Chen}
\end{center}

\begin{abstract}
	In this supplement, we provide the proofs for theoretical results of the paper.
\end{abstract}

\section{Proof of Theorem \ref{th:upper_bound_func}.} 

We prove Theorem \ref{th:upper_bound_func} by steps. Some key technical procedures are postponed to Lemmas  \ref{lm:hat_Sigma-to-hat_Sigma_0}, \ref{lm:factorization-lemma}, and \ref{lm:Hy-fan-norm}. 

\begin{enumerate}[leftmargin=*]
	\item[Step 1] Since we can always rescale the time domain, let $\mathcal{T} = [0, 1]$ throughout the proof without loss generality. We introduce some notations and prove basic properties in this step. Recall $T = \{t_1,\ldots, t_p\}$ is a regular grid on $\mathcal{T}$. Denote 
	\begin{equation}
	\begin{split}
	& \Sigma = \Cov((X(t_1), \ldots, X(t_p))^\top), \quad\Sigma_0 = \Cov((Z(t_1), \ldots, Z(t_p))^\top) \in \mathbb{R}^{p\times p}, \\
	& \Sigma_l = \Cov(X_{I_l}), \quad \Sigma_{0l} = \Cov(Z_{I_l})  \in \mathbb{R}^{|I_l|\times|I_l|},\quad l=1,\ldots, l_{\max}.
	\end{split}
	\end{equation} 
	Then
	$$\Sigma = \Sigma_0 + \sigma^2I_p, \quad \Sigma_{l} = \Sigma_{0l} + \sigma^2 I_{|I_l|},$$ 
	$\Sigma_l$ and $\Sigma_{0l}$ are submatrices of $\Sigma$ and $\Sigma_0$,
	\begin{equation}\label{eq:def-Sigma-01}
	\Sigma_l = \Sigma_{[I_l, I_l]},\quad \Sigma_{0l} = (\Sigma_0)_{[I_l, I_l]}, \quad l=1,\ldots, l_{\max}.
	\end{equation}
	For each subject $k$, recall $X_k(T) = (X_k(t_1), \ldots, X_k(t_p))^\top$ is the discretization of the sample path $X_k$. Given $G = G^{(r)} + G^{(-r)}$, we also decompose $\Sigma_0 = \Sigma_0^{(r)} + \Sigma_0^{-(r)}$, where $(\Sigma_0^{(r)})_{ij} = G^{(r)}(t_i, t_j), (\Sigma_0^{(-r)})_{ij} = G^{(-r)}(t_i, t_j)$. Suppose the eigenvalue decomposition of $\Sigma_0^{(r)}$ and $\Sigma_{0l}^{(r)}$ are
	\begin{equation}
	\Sigma_{0}^{(r)} = UDU^\top,\quad U\in \mathbb{O}_{p, r},\quad D \in\mathbb{R}^{r\times r} \text{ is diagonal};
	\end{equation}
	\begin{equation*}
	\Sigma_{0l}^{(r)} = (\Sigma_{0}^{(r)})_{[I_l, I_l]}, \quad \Sigma_{0l}^{(-r)} = (\Sigma_{0}^{(-r)})_{[I_l, I_l]}, \quad l=1,\ldots, l_{\max}.
	\end{equation*}
	Namely, $\Sigma_{0l}^{(r)}$ and $\Sigma_{0l}^{(-r)}$ are the submatrices of $\Sigma_{0}^{(r)}$ and $\Sigma_{0}^{(-r)}$. Then $\Sigma_{0l}^{(r)} + \Sigma_{0l}^{(-r)} = \Sigma_{0l}$ and $\Sigma_{0l}^{(r)} = U_{[I_l, :]}DU_{[I_l, :]}^\top$. It is also noteworthy that $\Sigma_{0l}^{(r)}$ and $\Sigma_{0l}^{(-r)}$ are not necessarily orthogonal, and $\Sigma_{0l}^{(r)}$ is not necessarily the best rank-$r$ approximation of $\Sigma_{0l}$. We also define
	\begin{equation}
	\begin{split}
	& A = UD^{1/2} \in \mathbb{R}^{p\times r}, \quad A_l = U_{[I_l, :]}D^{1/2} \in \mathbb{R}^{|I_l|\times r}, l=1,\ldots, l_{\max}\\
	& \text{then} \quad \Sigma_0^{(r)} = AA^\top, \quad \Sigma_{0l}^{(r)} = A_lA_l^\top.
	\end{split}
	\end{equation}
	Especially, $A$ and $A_l$ can be seen as the factors of $\Sigma_0$ and $\Sigma_{0l}$.
	
	Since $G(s_1, s_2)$ is Liptchitz, by Weyl's inequality \citep{weyl1949inequalities},
	\begin{equation}\label{ineq:liptchitz-1}
	\begin{split}
	& |\sigma_j(\Sigma_0)/p - \sigma_j(G)|\leq O(1/p), \quad \forall j; \quad \|\Sigma_0^{(-r)}\|_F/p \leq C\|G_0^{(-r)}\|_{HS} + O(1/p),\\
	\end{split}
	\end{equation}
	\begin{equation}\label{ineq:Sigma_0-spectral-norm}
	\|\Sigma_0\| \leq p\cdot \sigma_1(G) + O(1) \leq Cp,
	\end{equation}
	\begin{equation}\label{ineq:liptchitz-2}
	\|\Sigma_0^{(r)}\|_F \leq \|\Sigma_0\|_F \leq p\cdot \|G\|_{HS} + O(1) \leq Cp.
	\end{equation}
	We also have 
	\begin{equation}\label{eq:EX(t)^4}
	\begin{split}
	\mathbb{E}\|X(T)\|_2^4 = & \mathbb{E}\left(\sum_{i=1}^p X(T(i))^2 \right)^2 \overset{\text{Cauchy-Schwarz}}{\leq} p\sum_{i=1}^p \mathbb{E} X(T(i))^4\\
	\leq &  Cp^2\sup_t\mathbb{E}|X(t)|^4 \leq Cp^2.
	\end{split}
	\end{equation}
	Let $\mathcal{I}_l = [t_{(l-1)a+1} - 1/(2p), \{t_{(l-1)a+b} + 1/(2p)\}\wedge 1]$, then $\mathcal{I}_l$ is the time sub-domain corresponding to the grid indices subset $I_l$. By the construction of $I_l$ in \eqref{eq:I_l}, $I_l \cap I_{l+1} = \{la+1,\ldots, (l-1)a + b\}$, so $|I_l \cap I_{l+1}| = b-a$, $\mathcal{L}(\mathcal{I}_l \cap \mathcal{I}_{l+1}) \geq \frac{b-a}{p} \geq \kappa$ (introduced in Assumption \ref{as:sub-matrix}), thus 
	$$\sigma_r(G^{(r)}_{[\mathcal{I}_l \cap \mathcal{I}_{l+1}, \mathcal{I}_l \cap \mathcal{I}_{l+1}]})\geq c/\gamma \cdot \frac{\mathcal{L}(\mathcal{I}_l \cap \mathcal{I}_{l+1})}{\mathcal{L}(\mathcal{T})}$$ 
	based on the assumption. Provided that $p>C\gamma$ for large constant $C>0$, we further have
	\begin{equation}\label{ineq:sigma_r-Sigma_l}
	\begin{split}
	& \sigma_r\left(\Sigma^{(r)}_{0, [I_l\cap I_{l+1}, I_l \cap I_{l+1}]}\right) = \sigma_r\left(G^{(r)}_{[\mathcal{I}_l \cap \mathcal{I}_{l+1}, \mathcal{I}_l \cap \mathcal{I}_{l+1}]}\right)\cdot p + O(1)\\
	\overset{\text{Assumption 2}}{\geq} & tr(G) p/\gamma \cdot \frac{\mathcal{L}(\mathcal{I}_l\cap \mathcal{I}_{l+1})}{\mathcal{L}(\mathcal{T})} + O(1) \geq cp/\gamma. 
	\end{split}
	\end{equation}
	The constant $c$ here may depend on constant $\kappa$. Provided that $\|D\|/p = \sigma_1(\Sigma_0)/p \leq \sigma_1(G) + O(1/p) \leq C$, $A_{[I_l\cap I_{l+1}, :]}A_{[I_l\cap I_{l+1}, :]}^\top = \Sigma^{(r)}_{0, [I_l\cap I_{l+1}, I_l\cap I_{l+1}]}$, we further have
	\begin{equation}\label{ineq:sigma_rA_I_l-l+1}
	\sigma_r\left(A_{[I_l\cap I_{l+1}, :]}\right) = \sqrt{\sigma_r\left(\Sigma^{(r)}_{0, [I_l\cap I_{l+1}, I_l\cap I_{l+1}]}\right)} \geq c\sqrt{p/\gamma}, \quad l=1,\ldots, l_{\max}.
	\end{equation}
	\begin{equation}
	\|A\| \leq \sqrt{\sigma_1(\Sigma_0)} \leq C\sqrt{p}.
	\end{equation}

	\item[Step 2] Our aim in this step is to develop a perturbation bound for $\hat{A}_l$, i.e. to characterize the distance between $\hat{A}_l$ and $A_l$ for each $l=1,\ldots, l_{\max}$. Recall $\Delta_l = \|\hat{\Sigma}_l - \Sigma_l\|_F$, $\lambda = \|\Sigma_0^{(-r)}\|_F$. By Lemma \ref{lm:hat_Sigma-to-hat_Sigma_0} and $b \geq 2r$,
	\begin{equation}\label{ineq:thm1-intermediate-1}
	\begin{split}
	\left\|\hat{A}_l\hat{A}_l^\top - \Sigma_{0l}^{(r)}\right\|_F = \left\|\hat{A}_l\hat{A}_l^\top - A_lA_l^\top \right\|_F \leq C\left(\Delta_l + \|\Sigma_{0}^{(-r)}\|_F\right), l=1,\ldots, l_{\max}.
	\end{split}
	\end{equation}
	By Lemma \ref{lm:factorization-lemma}, there exists $Q_l \in \mathbb{O}_{r}$ such that
	\begin{equation}\label{ineq:th1-intermediate}
	\left\|\hat{A}_l - A_lQ_l \right\|_F \leq \frac{\|\hat{A}_l\hat{A}_l^\top - A_lA_l^\top\|_F}{\sqrt{\sigma_r(A_l)\sigma_r(\hat{A}_l)}} \leq \frac{C\left(\Delta_l + \lambda\right)}{\sqrt{\sigma_r(A_l)\sigma_r(\hat{A}_l)}} \wedge \sqrt{\|\hat{A}\|_F^2 + \|A\|_F^2}.
	\end{equation}
	We analyze each term in \eqref{ineq:th1-intermediate} as follows. By \eqref{ineq:sigma_rA_I_l-l+1}, 
	$$\sigma_r(A_l) \geq c\sqrt{p/\gamma};$$
	\begin{equation}\label{ineq:sigma-r-hat-A-l}
	\begin{split}
	& \sigma_r^2(\hat{A}_l) = \sigma_r(\hat{A}_l\hat{A}_l^\top) \overset{\eqref{ineq:thm1-intermediate-1}}{\geq} \sigma_r(\Sigma_{0l}^{(r)}) - C\left(\Delta_l + \lambda\right)\\
	\overset{\eqref{ineq:sigma_r-Sigma_l}}{\geq} & cp/\gamma - C\left(\Delta_l +\lambda\right).
	\end{split}
	\end{equation}
	\begin{equation*}
	\begin{split}
	\|A_l\|_F^2 = & \tr(A_lA_l^\top) = \sum_{j=1}^r \sigma_j(\Sigma_{0l}^{(r)}) \leq \sqrt{r} \|\Sigma_{0l}^{(r)}\|_F \leq \sqrt{r}\|\Sigma_0^{(r)}\|_F\\
	\overset{\eqref{ineq:liptchitz-2}}{\leq} & \sqrt{r} (p\cdot \|G\|_{HS} + O(1)) \leq Cp\sqrt{r},
	\end{split}
	\end{equation*}
	\begin{equation*}
	\begin{split}
	\|\hat{A}_l\|_F^2 = & \tr(\hat{A}_l\hat{A}_l^\top) \leq  \sqrt{r}\|\hat{A}_l\hat{A}_l^\top\|_F \overset{\eqref{ineq:thm1-intermediate-1}}{\leq} \sqrt{r}\|\Sigma_{0l}\|_F +  C\sqrt{r}\left(p + \Delta_l+\lambda\right)\\
	\leq & C\sqrt{r}\left(p + \Delta_l + \lambda\right).
	\end{split}
	\end{equation*}
	By combining the previous inequalities, we conclude that 
	\begin{equation}\label{ineq:step1-conclusion}
	\begin{split}
	\left\|\hat{A}_{l} - A_{l} Q_{l} \right\|_F \leq & \frac{C(\Delta_l + \lambda)}{\left(\frac{p}{\gamma}\left(\frac{p}{\gamma} - C(\Delta_l+\lambda)\right)_+\right)^{1/4}}\wedge C\left\{r^{1/4}(p^{1/2}+\lambda^{1/2}+\Delta_l^{1/2})\right\}\\ 
	:= & \tilde{\Delta}_l,
	\end{split}
	\end{equation}
	for $ l = 1,\ldots, l_{\max}$ and some uniform constant $C>0$. Here $(x)_+ = \max\{x, 0\}$ for any $x\in \mathbb{R}$.
	\item[Step 3] In this step, we assume \eqref{ineq:step1-conclusion} hold. Recall $\hat{O}_l$ is calculated sequentially. In this step, we study how the statistical error of $\hat{O}_l$ is accumulated based on \eqref{ineq:step1-conclusion} in this step. Ideally speaking, $\hat{O}_{l+1}$ can be seen as an estimation of $(Q_{l+1}^\top Q_1\hat{O}_1)$. Specifically, we aim to show that there exists a uniform constant $C>0$ such that 
	\begin{equation}\label{ineq:induction}
	\left\|Q_{l+1} \hat{O}_{l+1}  - Q_1 \hat{O}_1 \right\|_F  \leq \frac{C\tilde{\Delta}}{\sqrt{p/\gamma}}, \quad l=0,\ldots, l_{\max}-1.
	\end{equation}
	and 
	\begin{equation}\label{ineq:A_lO_l-A_lO_1}
	\left\|\hat{A}_l \hat{O}_l - A_l Q_1 \hat{O}_1\right\|_F \leq C\sqrt{\gamma}\tilde{\Delta}, \quad l=1,\ldots, l_{\max}.
	\end{equation}
	Here, $\tilde{\Delta} = \sum_{l=1}^{l_{\max}} \tilde{\Delta}_l$. First, for each $l=1,\ldots, l_{\max}-1$, we introduce
	\begin{equation*}
	B_l^{(2)} := (\hat{A}_{l})_{[(a+1):b, :]}\cdot \hat{O}_l \in \mathbb{R}^{(b-a) \times r},\quad B_{l+1}^{(1)} := (\hat{A}_{l+1})_{[1:(b-a), :]} \in \mathbb{R}^{(b-a) \times r}.
	\end{equation*}
	Essentially, $B_l^{(2)}$ contains the last $(b-a)$ rows of $\hat{A}_l$ after rotation and $B_{l+1}^{(1)}$ contains the first $(b-a)$ rows of $\hat{A}_{l+1}$ before rotation. According to the proposed procedure \eqref{eq:hat_O_l}, 
	\begin{equation}\label{ineq:th1-step2-1}
	\hat{O}_{l+1} = \argmin_{O\in \mathbb{O}_r}\left\|B_l^{(2)} - B_{l+1}^{(1)}\cdot O\right\|_F.
	\end{equation}
	Since $B_l^{(2)}$ and $B_{l+1}^{(1)}$ are submatrices of $\hat{A}_l$ and $\hat{A}_{l+1}$ respectively, they also satisfy
	\begin{equation}\label{ineq:th1-step2-2}
	\begin{split}
	& \left\|B_l^{(2)} - A_{l,[(a+1):b, :]} Q_l\hat{O}_l \right\|_F = \left\|\hat{A}_{l, [(a+1):b, :]}\hat{O}_l - A_{l, [(a+1):b, :]}Q_l\hat{O}_l\right\|_F\\
	\leq & \left\|\hat{A}_l\hat{O}_l - A_l Q_l \hat{O}_l\right\|_F = \left\|\hat{A}_l - A_l Q_l\right\|_F \overset{\eqref{ineq:step1-conclusion}}{\leq} \tilde{\Delta}_l.
	\end{split}
	\end{equation}
	\begin{equation}\label{ineq:th1-step2-3}
	\begin{split}
	& \left\|B_{l+1}^{(1)} - A_{l+1,[1:(b-a), :]} Q_{l+1} \right\|_F = \left\|\hat{A}_{l+1, [1:(b-a), :]} - A_{l+1, [1:(b-a), :]}Q_{l+1}\right\|_F\\
	\leq & \left\|\hat{A}_{l+1} - A_{l+1} Q_{l+1}\right\|_F \overset{\eqref{ineq:step1-conclusion}}{\leq} \tilde{\Delta}_{l+1}.
	\end{split}
	\end{equation}
	More importantly, $A_{l,[(a+1):b, :]} = A_{l+1,[1:(b-a), :]} = A_{[I_l\cap I_{l+1}, :]}$, as they actually represent the same submatrix of $A$. Then \eqref{ineq:th1-step2-1}--\eqref{ineq:th1-step2-3} and Lemma \ref{lm:rotations-Wahba} yield 
	\begin{equation}\label{ineq:th1-step2-4}
	\begin{split}
	& \left\|\hat{O}_{l+1} - Q_{l+1}^\top Q_l\hat{O}_l\right\|_F  = \left\|Q_{l+1}\hat{O}_{l+1} - Q_l\hat{O}_l\right\|_F\\
	\leq & \frac{2\left(\left\|B_l^{(2)} - A_{l,[(a+1):b, :]} Q_l\hat{O}_l \right\|_F + \left\|B_{l+1}^{(1)} - A_{l+1,[1:(b-a), :]} Q_{l+1} \right\|_F\right)}{\sigma_r(A_{I_l\cap I_{l+1}})}\\
	\leq & \frac{2(\tilde{\Delta}_l + \tilde{\Delta}_{l+1})}{\sigma_r(A_{[I_l\cap I_{l+1}, :]})} \overset{\eqref{ineq:sigma_rA_I_l-l+1}}{\leq} \frac{C(\tilde{\Delta}_l + \tilde{\Delta}_{l+1})}{\sqrt{p/\gamma}}.
	\end{split}
	\end{equation}
	Recall $\tilde{\Delta} = \sum_{l=1}^{l_{\max}} \tilde{\Delta}_l$. Thus, 
	\begin{equation*}
	\begin{split}
	& \left\|Q_{l} \hat{O}_{l}  - Q_1 \hat{O}_1 \right\|_F  \leq \sum_{k=1}^{l-1}\left\|Q_{k+1}\hat{O}_{k+1} - Q_k\hat{O}_k\right\|_F \\
	\leq & \sum_{k=1}^{l-1} \frac{C(\tilde{\Delta}_l+\tilde{\Delta}_{l+1})}{\sqrt{p/\gamma}}= \frac{C\tilde{\Delta}}{\sqrt{p/\gamma}},
	\end{split}
	\end{equation*}
	which has finished the proof for \eqref{ineq:induction}. Then
	\begin{equation}
	\begin{split}
	& \left\|\hat{A}_l\hat{O}_l - A_l Q_1\hat{O}_1\right\|_F \leq \left\|\hat{A}_l\hat{O}_l - A_l Q_l\hat{O}_l\right\|_F + \left\|A_l Q_l\hat{O}_l - A_l Q_1\hat{O}_1\right\|_F\\
	\leq & \left\|\hat{A}_l - A_l Q_l \right\|_F + \left\|A_l\right\|\cdot \left\|Q_l\hat{O}_l - Q_1\hat{O}_1\right\|_F\\
	\overset{\eqref{ineq:step1-conclusion}\eqref{ineq:induction}}{\leq} & \tilde{\Delta}_l +   \|\Sigma_{0l}\|^{1/2} \cdot\frac{C\tilde\Delta}{\sqrt{p/\gamma}} \overset{\eqref{ineq:Sigma_0-spectral-norm}}{\leq} C\sqrt{\gamma}\tilde{\Delta}
	\end{split}
	\end{equation}
	for $l=1,\ldots, l_{\max}$, which has finished the proof for \eqref{ineq:A_lO_l-A_lO_1}.
	
	\item[Step 4] In this step, we develop the error bound from sequential aggregation based on \eqref{ineq:A_lO_l-A_lO_1}. Recall
	\begin{equation}
	\tilde{A}\in \mathbb{R}^{p\times r}, \quad \tilde{A}_{[i, :]} = \frac{\sum_{l: i\in I_l}\hat{A}_{l, [i, :]}^\ast}{\left|\left\{l: i\in I_l\right\}\right|}.
	\end{equation}
	The direct way to analyze $\tilde{A}$ is complicated. We instead consider the following half integers between $1/2$ and $p+1/2$, 
	\begin{equation}\label{eq:th_1-bar}
	\begin{split}
	\mathcal{B} = & \left\{.5, a+.5, \ldots, (l_{\max}-1)a + .5\right\}\\
	& \cup \left\{b+.5, b+a+.5, \ldots, b+(l_{\max}-1)a + .5\right\},
	\end{split}
	\end{equation}
	and divide the whole index set $\{1,\ldots, p\}$ into pieces, say $K_1,\ldots, K_m$, by inserting ``bars" with the half integers in \eqref{eq:th_1-bar}. For example, when $p = 10, b = 5, a=3$, then $\mathcal{B} = \{.5, 3.5, 6.5\} \cup \{5.5, 8.5, 11.5\}$, and $\{1,\ldots, 10\}$ is divided as the following subsets	
	$$K_1,\ldots, K_5 \quad = \quad \{1,2,3\},\quad \{4, 5\}, \quad\{6\}, \quad \{7, 8\}, \quad \{9, 10\}.$$
	Such a division has two important properties,
	\begin{itemize}
		\item Given $|\mathcal{B}| \leq 2l_{\max}$ and $0.5\in \mathbb{B}$, $\{1,\ldots, p\}$ is divided into at most $2l_{\max}$ intervals, so $m\leq 2l_{\max}$.
		\item For any piece $K_s$ and two indices $i, j \in K_s$, we must have
		$$\{l: i \in I_l\}  = \{l: j\in I_l\},$$
		namely Indices $i$ and $j$ belong to the same set of sub-intervals $\{I_l\}$. Thus, we can further denote $J_{K_s} = \{l: i \in I_l, \forall i \in K_s\}$ as the sub-intervals that covers $K_s$. Then the following equality holds,
		\begin{equation}\label{eq:th1_step3-1}
		\tilde{A}_{[K_s, :]} = \frac{1}{|J_{K_s}|}\sum_{l\in J_{K_s}} \hat{A}_{l, [K_s, ]}^\ast .
		\end{equation}
		Based on the definition of $J_{K_s}$, we also know
		\begin{equation}\label{eq:K_s-subset-I_l}
		\forall l\in J_{K_s}, \quad K_s\subseteq I_l.
		\end{equation}
	\end{itemize}
	Based on these two points, we analyze $\tilde{A}$ on each piece $K_s$ and then aggregate as follows,
	\begin{equation}\label{ineq:th1-step3-1}
	\begin{split}
	\left\|\tilde{A} - A Q_1\hat{O}_1\right\|_F = & \sqrt{\sum_{s=1}^m \left\|\tilde{A}_{[K_s, :]} - A_{[K_s, :]} Q_1\hat{O}_1\right\|_F^2}\\
	\overset{\eqref{eq:th1_step3-1}}{=} & \sqrt{\sum_{s=1}^m \left\|\frac{1}{|J_{K_s}|} \sum_{l \in J_{K_s}} \left(\hat{A}_{l, [K_s, :]}^\ast - A_{l, [K_s, :]}Q_1\hat{O}_1\right)\right\|_F^2}.\\
	\end{split}
	\end{equation}
	Now for each $s = 1,\ldots, m$, 
	\begin{equation}\label{ineq:th1-step3-2}
	\begin{split}
	& \left\|\frac{1}{|J_{K_s}|} \sum_{l \in J_{K_s}} \left(\hat{A}_{l, [K_s, :]}^\ast - A_{[K_s, :]}Q_1\hat{O}_1\right)\right\|_F \\
	\leq & \frac{1}{|J_{K_s}|} \sum_{l\in J_{K_s}}\left\|\left(\hat{A}_{l, [K_s, :]}^\ast - A_{[K_s, :]}Q_1\hat{O}_1\right)\right\|_F\\
	\overset{\eqref{eq:K_s-subset-I_l}}{\leq} & \frac{1}{|J_{K_s}|} \sum_{l\in J_{K_s}}\left\|\left(\hat{A}_{l}\hat{O}_l - A_{l}Q_1\hat{O}_1\right)\right\|_F \overset{\eqref{ineq:A_lO_l-A_lO_1}}{\leq} C\gamma^{1/2}\tilde{\Delta}.
	\end{split}
	\end{equation}
	Combining \eqref{ineq:th1-step3-1} and \eqref{ineq:th1-step3-2}, we obtain
	\begin{equation}\label{ineq:th1-step3-3}
	\left\|\tilde{A} - AQ_1\hat{O}_1\right\|_F \leq C\sqrt{m} \gamma^{1/2}\tilde{\Delta} \leq Cl_{\max}\gamma^{1/2}\tilde{\Delta}.
	\end{equation}
	By definition of $l_{\max}$, $b/p \leq Cl_{\max}^{-1}$, thus $l_{\max}\leq C$. Provided that $\sigma_{\max}(A) = \|\Sigma_0\|$ and $n^\ast \geq C \gamma^2$ for large constant $C$,
	$$\sigma_{\max}(\tilde{A}) \leq \sigma_{\max}(AQ_1\hat{O}_1) + \|\tilde{A} - AQ_1\hat{O}_1\|_F \overset{\eqref{ineq:th1-step3-3}}{\leq} Cp^{1/2} + C\gamma^{1/2}\tilde{\Delta}.$$
	Then the following inequality holds,
	\begin{equation}
	\begin{split}
	& \left\|\hat{\Sigma}_0 - \Sigma_0^{(r)}\right\|_F = \left\|\tilde{A}\tilde{A}^\top - AA^\top \right\|_F \\
	\leq & \left\|\tilde{A}\tilde{A}^\top - AQ_1\hat{O}_1\tilde{A}^\top\right\|_F + \left\|AQ_1\hat{O}_1\tilde{A}^\top - AQ_1\hat{O}_1\hat{O}_1^\top Q_1^\top A^\top \right\|_F\\
	\overset{\eqref{ineq:th1-step3-3}}{\leq} & \sigma_{\max}(\tilde{A}^\top) \cdot C\gamma^{1/2}\tilde{\Delta} + \sigma_{\max}(A) \cdot C\gamma^{1/2}\tilde{\Delta}\\
	\leq & C\gamma^{1/2}\tilde{\Delta}\left(Cp^{1/2}+C\gamma^{1/2}\tilde{\Delta}\right).
	\end{split}
	\end{equation}
	Given $\Sigma_0 = \Sigma_0^{(r)} + \Sigma_0^{(-r)}$ and $\|\Sigma_0^{(-r)}\|_F\leq \lambda$, 
	in summary, we have proved the upper bound
	$$\left\|\hat{\Sigma}_0 - \Sigma_0\right\|_F \leq C\gamma^{1/2}\tilde{\Delta}\left(Cp^{1/2}+C\gamma^{1/2}\tilde{\Delta}\right) + \lambda = C(\gamma p)^{1/2}\tilde{\Delta} + C\gamma \tilde{\Delta}^2 + \lambda.$$
	\item[Step 5] It remains to develop the expected error upper bound for $\hat{\Sigma}_0$ and $\hat{G}$. Recall $\Cov(X_{I_l}) = \Sigma_l$. If $\hat{\Sigma}_l$ is calculated from complete samples by \eqref{eq:hat_Sigma_complete} in Step 2, we have
	\begin{equation}\label{ineq:EDelta_l}
	\begin{split}
	\mathbb{E}\Delta_l^2 = & \mathbb{E}\left\|\frac{1}{n_l^\ast}\sum_{k\in J_l}((X_k)_{I_l} - \bar{X}_{I_l})((X_k)_{I_l} - \bar{X}_{I_l})^\top - \Sigma_l\right\|_F^2 \\
	\leq & \frac{C}{n_l^\ast} \mathbb{E} \left\|((X_k)_{I_l} - \mu(I_l))((X_k)_{I_l} - \mu(I_l))^\top - \Sigma_l\right\|_F^2 \\
	\leq & \frac{C}{n_l^\ast} \mathbb{E} \|(X_k)_{I_l}(X_k)_{I_l}^\top\|_F^2 \leq \frac{1}{n_l^\ast}\mathbb{E}\|X(T)\|_2^4 \overset{\eqref{eq:EX(t)^4}}{\leq} \frac{Cp^2}{n_l^\ast}.
	\end{split}
	\end{equation}
	Under the incomplete observation scenario (Step 2'), we have
	\begin{equation}\label{ineq:EDelta_l-incomplete}
	\begin{split}
	& \mathbb{E}\Delta_l^2 \\
	= & \sum_{i, j =1}^{|I_l|}\mathbb{E} \Bigg\{\frac{\sum_{k: I_l(i), I_l(j) \in T_k} \left(X_k(I_l(i)) - \bar{X}(I_l(i))\right)\left(X_k(I_l(j)) - \bar{X}(I_l(j))\right)}{(n^\ast)_{i,j,l}} \\
	& \qquad - \Sigma_{l, ij}\Bigg\}\\
	= & \sum_{i,j =1}^{|I_l|} \frac{\mathbb{E}\left\{\left(X(I_l(i)) - \mu(I_l(i))\right)\left(X(I_l(j)) - \mu(I_l(j))\right) - \Sigma_{l, ij}\right\}^2}{(n^\ast)_{i,j,l}}\\
	\leq & C\sum_{i,j =1}^{|I_l|} \frac{\mathbb{E}\left\{X(I_l(i)) \cdot X(I_l(j))\right\}^2}{(n^\ast)_{i,j,l}} \leq C\sum_{i, j=1}^{|I_l|}\frac{\mathbb{E}X(I_l(i))^4 + \mathbb{E}X(I_l(j))^4}{2n^\ast_{i,j,l}}\\
	\leq & \frac{C|I_l|^2}{n^\ast} \sup_t \mathbb{E}X(t)^4 \leq \frac{Cp^2}{n^\ast}.
	\end{split}
	\end{equation}
	Now we analyze $\|\hat{\Sigma}_0 - \Sigma_0\|_F$ in two scenarios under the complete sample case (Step 2). The incomplete sample case (Step 2') similarly follows. Recall the definitions of $\tilde{\Delta}_l$ and $\tilde{\Delta}$,
	$$\tilde{\Delta}_l = \frac{C(\Delta_l + \lambda)}{\left(\frac{p}{\gamma}\left(\frac{p}{\gamma} - C(\Delta_l+\lambda)\right)_+\right)^{1/4}}\wedge C\left\{r^{1/4}(p^{1/2}+\lambda^{1/2}+\Delta_l^{1/2})\right\}, \tilde{\Delta} = \sum_l \tilde{\Delta}_l. $$
	Let 
	\begin{equation}\label{eq:B-good-event}
	\begin{split}
	B = & \left\{p/\gamma - C(\Delta_l + \lambda)\geq p/(2\gamma), \forall l=1,\ldots, l_{\max}\right\}\\
	= & \left\{C(\Delta_l + \lambda)\leq p/(2\gamma), \forall l=1,\ldots, l_{\max}\right\}
	\end{split}
	\end{equation}
	be a ``good" event. By Markov's inequality, 
	\begin{equation}\label{ineq:B^c-chance}
	\mathbb{P}(B^c) \leq \sum_{l=1}^{l_{\max}}\frac{\mathbb{E}\{C(\Delta_l+\lambda)\}^2}{p^2/(2\gamma)^2} \overset{\eqref{ineq:EDelta_l}\eqref{ineq:EDelta_l-incomplete}}{\leq} \frac{Cl_{\max} p^2/ n^\ast}{p^2/\gamma^2} \leq C\gamma^2/n^\ast.
	\end{equation} 
	When $B$ holds, note that $n^\ast \geq Cp$, we have
	\begin{equation*}
	\begin{split}
	& \mathbb{E}\left\|\hat{\Sigma}_0 - \Sigma_0\right\|_F 1_B \leq \mathbb{E}\left\{C(\gamma p)^{1/2}\tilde{\Delta} + Cr\tilde{\Delta}^2 + \lambda \right\}1_B \\
	\leq & C(\gamma p)^{1/2}\sum_{l=1}^{l_{\max}}\mathbb{E}\frac{(\Delta_l + \lambda)}{(p/\gamma)^{1/2}} + C\gamma \cdot \mathbb{E}\frac{\{\sum_{l=1}^{l_{\max}}(\Delta_l+\lambda)\}^2}{(p/\gamma)} + \lambda\\
	\leq & C\gamma \sqrt{p^2/n^\ast} + C\gamma p/n^\ast \leq Cp\sqrt{\gamma^2/n^\ast}.
	\end{split}
	\end{equation*}
	When $B^c$ holds, given $n^\ast \geq \gamma^2, p\geq \gamma$, we have
	\begin{equation*}
	\begin{split}
	& \mathbb{E}\left\|\hat{\Sigma}_0 - \Sigma_0\right\|_F1_{B^c} \leq \mathbb{E}C\left\{\gamma^{1/4}(p^{1/2}+\lambda^{1/2}+\sum_{l=1}^{l_{\max}}\Delta_l^{1/2})\right\} 1_{B^c} \\
	\leq & CP(B^c) \gamma^{1/4}p^{1/2} + \sum_{l=1}^{l_{\max}} C\gamma^{1/4} \left(\mathbb{E} (\Delta_l^{1/2})^2\right)^{1/2}\cdot \left(\mathbb{E} 1_{B}^2\right)^{1/2}\\
	\leq & \frac{C\gamma^{9/4}p^{1/2}}{n^\ast} + Cl_{\max}\gamma^{1/4}\left(p/(n^\ast)^{1/2}\cdot \gamma^2/n^\ast\right)^{1/2} \\
	\leq & \frac{C\gamma^{9/4}p^{1/2}}{n^\ast} + \frac{C\gamma^{5/4}p^{1/2}}{(n^\ast)^{3/4}}\leq Cp\sqrt{\gamma^2/n^\ast}.
	\end{split}
	\end{equation*}
	In summary,
	\begin{equation*}
	\mathbb{E}\left\|\hat{\Sigma}_0 - \Sigma_0\right\|_F \leq Cp\sqrt{\gamma^2/n^\ast}.
	\end{equation*}
	
	Finally, since $\Sigma_0$ is a $p$-by-$p$ linear interpolation for $G$, we finally have
	\begin{equation*}
	\left\|\hat{G} - G\right\|_{HS} \leq \frac{1}{p} \|\hat{\Sigma}_0 - \Sigma_0\|_F + O(p^{-1}) = O(\sqrt{\gamma^2/n^\ast} + p^{-1}).
	\end{equation*}
	\quad $\square$
\end{enumerate}

\section{Proof of Proposition \ref{pr:complete}}

The key of developing a sharper rate for $\hat{\Sigma}$ is on a better estimation bound for $\|\hat{A}_l-A_lO\|_F$, where $\hat{A}_l$ is the estimated factor computed in Step 3 of the proposed procedure. The essence of the sharper bound relies on the following lemma.
\begin{Lemma}\label{lm:factor-estimation}
	Suppose all conditions in Theorem \ref{th:upper_bound_func} and Proposition \ref{pr:complete} hold. Recall $\hat{A}_l$ is the estimation of the factor of each piece calculated in Step 3 in the proposed procedure. Then there exists a ``good event" $B_\ast$ (defined later in Equation \ref{eq:Bast-good-event}) that happens with probability at least $1 - C\gamma r/n^\ast$, such that
	$$\mathbb{E}\min_{O\in \mathbb{O}_r}\|\hat{A}_l - A_lO\|_F^2 \cdot 1_{\{B^\ast \text{ holds}\}} \leq Cpr/n_l^\ast, \quad \forall l=1,\dots, l_{\max}.$$	
\end{Lemma}

{\bf\noindent Proof of Lemma \ref{lm:factor-estimation}.} We assume $\mu(t)=0$ without changing the covariance estimators essentially. Note that the sample covariance $\hat{\Sigma}_l$ is calculated in Step 2 as 
$$\hat{\Sigma}_l = \frac{1}{n_l^\ast} \sum_{k\in J_l} \left((X_k)_{I_l} - (\bar{X})_{I_l}\right)\left((X_k)_{I_l} - \bar{X}_{I_l}\right)^\top.$$ 
The proof of this lemma is divided into steps.
\begin{enumerate}[leftmargin=*]
	\item[Step 1] We introduce a series of notation in addition to the symbols in the proof of Theorem \ref{th:upper_bound_func} here. Based on Karhunen-Lo\`eve decomposition, the continuous sample trajectory $X_k(t) = Z_k(t) + \epsilon_k(t)$ can be decomposed into three parts: the leading part of signal, the non-leading part of signal, and the noise:
	\begin{equation}\label{eq:X-k-observation}
	X_k(t) = \sum_{j=1}^r \xi_{jk}\phi_j(t)  + \sum_{j \geq r+1} \xi_{jk}\phi_j(t)  + \epsilon_{k}(t), \quad k\in J_l.
	\end{equation}
	Then, $\lambda_j(G) = \Var(\xi_{kj})$. Let $\bar{\xi}_{jk} = \xi_{jk}/\lambda^{1/2}_j(G)$ be the normalized score. We further define
	\begin{equation}
	S = \begin{bmatrix}
	\bar{\xi}_{11} & \cdots & \bar{\xi}_{1n}\\
	\vdots & & \vdots \\
	\bar{\xi}_{r1} & \cdots & \bar{\xi}_{rn}
	\end{bmatrix} \in \mathbb{R}^{r\times n},
	\end{equation}
	\begin{equation*}
	\Phi_l = \begin{bmatrix}
	(\phi_1)_{I_l(1)} \cdot \lambda_1^{1/2}(G) & \cdots  & (\phi_1)_{I_l(1)} \cdot \lambda_r^{1/2}(G)\\
	\vdots & & \vdots\\
	(\phi_1)_{I_l(|I_l|)} \cdot \lambda_1^{1/2}(G) & \cdots  & (\phi_1)_{I_l(|I_l|)} \cdot \lambda_r^{1/2}(G)
	\end{bmatrix}\in \mathbb{R}^{|I_l| \times r}
	\end{equation*}
	as the matrix of leading scores and the discretized loadings, respectively. Then, $\Phi_l$ matches the definition \eqref{eq:def-Sigma-01} in Theorem \ref{th:upper_bound_func} as
	\begin{equation}\label{eq:PhiPhi^top}
	\Phi_l\Phi_l^\top = \Sigma_{0l}^{(r)} \in \mathbb{R}^{|I_l|\times|I_l|}.
	\end{equation} 
	We further let $Z^{(-r)}_k(t) = \sum_{j \geq r+1} \xi_{jk}\phi_j(t) $ be the tail part of sample. By restricting \eqref{eq:X-k-observation} onto the index set $I_l$, one has
	\begin{equation*}
	\begin{split}
	(X_k)_{I_l} = & \sum_{j=1}^r \xi_{jk}(\phi_{j})_{I_l}  + \sum_{j \geq r+1} \xi_{jk}(\phi_j)_{I_l}  + (\epsilon_{k})_{I_l}\\
	= & \Phi_l S_{[:, k]} + (Z_{k}^{(-r)})_{I_l} + (\epsilon_k)_{I_l}, \quad k\in J_l.
	\end{split}
	\end{equation*}
	Based on the proof of Theorem \ref{th:upper_bound_func}, $\rank(\Sigma_{0l}^{(r)}) = r$ and 
	$$\sigma_j^2(\Phi) \overset{\eqref{eq:PhiPhi^top}}{=} \sigma_j(\Sigma^{(r)}_{01}) \overset{\eqref{ineq:sigma_r-Sigma_l}}{=} p\sigma_j(G^{(r)}_{[\mathcal{I}_l, \mathcal{I}_l]}) + O(1), \quad j=1,\ldots, r,$$
	\begin{equation}\label{ineq:Phi-F}
	\begin{split}
	\|\Phi\|_F = (\tr(\Phi\Phi^\top))^{1/2} = & \left(\tr(\Sigma_{0l}^{(r)})\right)^{1/2}\leq \left(p\cdot \tr(G^{(r)}_{[\mathcal{I}_l, \mathcal{I}_l]}) + O(r)\right)^{1/2} \\
	\overset{\text{Assumption \ref{as:sub-matrix}}}{\leq} &  Cp^{1/2}.
	\end{split}
	\end{equation}
	In particularly,
	\begin{equation}\label{eq:sigma_r-Phi}
	\sigma_r(\Phi) = \sigma_r(A_l) = \sqrt{\sigma_r(\Sigma_{0, [I_l, I_l]}^{(r)})} \overset{\eqref{ineq:sigma_r-Sigma_l}}{\geq} c\sqrt{p/\gamma}.
	\end{equation}
	Recall the central goal of this proposition is to provide an upper bound for $\min_{O\in \mathbb{O}_r}\|\hat{A}_l - A_lO\|_F$. One can only show $\min_O \|\hat{A}_l - A_lO\|_F^2 \leq Cp\gamma/n_l^\ast$ by directly applying Lemma \ref{lm:factorization-lemma} on $\hat{\Sigma}_{0l}$ and $\Sigma_{0l}^{(r)}$. Instead, we introduce a ``bridge" covariance in this proof
	\begin{equation}
	\begin{split}
	\bar{\Sigma}_{0l} = & \frac{1}{n_l^\ast} \sum_{k\in J_l}(\Phi_l S_{[:, k]})(\Phi_l S_{[:, k]})^\top  = \frac{1}{n_l^\ast} \Phi_l S_{J_l} S_{J_l}^\top \Phi_l^\top \in \mathbb{R}^{|I_l|\times |I_l|}. 
	\end{split}
	\end{equation}
	Let $\bar{A}_l = \Phi_l S_{J_l}/\sqrt{n_l^\ast}$. Then for all $Q \in \mathbb{O}_r$, we have
	\begin{equation*}
	\begin{split}
	\min_{O\in \mathbb{O}_r} \|\hat{A}_l - A_l O\|_F \leq & \min_{O\in \mathbb{O}_r} \left\{\|\hat{A}_l - \bar{A}_lQ\|_F + \|\bar{A}_lQ - A_lO\|_F\right\}\\
	= & \|\hat{A}_l - \bar{A}_lQ\|_F + \min_{O\in \mathbb{O}_r}\|\bar{A}_l - A_lOQ^\top\|_F. 
	\end{split}
	\end{equation*}
	By taking the infimum over $Q \in \mathbb{O}_r$, we obtain the following triangle inequality,
	\begin{equation}\label{ineq:triangle}
	\min_{O\in \mathbb{O}_r} \|\hat{A}_l - A_l O\|_F \leq \min_{O\in \mathbb{O}_r}\|\hat{A}_l - \bar{A}_lO\|_F + \min_{O\in \mathbb{O}_r}\|\bar{A}_l - A_lO\|_F.
	\end{equation}
	In the next two steps, we give upper bounds for $\min_{O\in \mathbb{O}_r} \|\bar{A}_l - A_l O\|_F$ and $\min_{O\in \mathbb{O}_r} \|\hat{A}_l - \bar{A}_l O\|_F$, respectively.
	\item[Step 2] Since $\rank(S_{J_l}S_{J_l}^\top/n_l^\ast) = r$, we can further factorize 
	$$S_{J_l}S_{J_l}^\top/n_l^\ast = F_lF_l^\top$$ 
	for some $F_l\in \mathbb{R}^{r\times r}$. Then,
	\begin{equation*}
	\begin{split}
	& \sigma_{\min}\left(S_{J_l}/\sqrt{n_l^\ast}\right) = \sqrt{\sigma_{\min}\left(S_{J_l}S_{J_l}^\top/n_l^\ast\right)} = \sigma_{\min}(F_l),\\
	&  \sigma_{\max}\left(S_{J_l}/\sqrt{n_l^\ast}\right) = \sqrt{\sigma_{\max}\left(S_{J_l}S_{J_l}^\top/n_l^\ast\right)} = \sigma_{\max}(F_l).
	\end{split}
	\end{equation*} 
	Suppose $$F = U_F\Sigma_F V_F^\top, \quad U_F, V_F\in \mathbb{O}_r, \quad \Sigma_F\in \mathbb{R}^{r\times r}$$ 
	is the singular value decomposition. Since $\Sigma_F$ is diagonal, we have
	\begin{equation}\label{ineq:Sigma_F-I}
	\begin{split}
	\|\Sigma_F - I_{r\times r}\| \leq & \max\left\{\sigma_{\max}(F_l) - 1, 1 - \sigma_{\min}(F_l)\right\}\\
	= & \max\left\{\sigma_{\max}(S_{J_l}/\sqrt{n_l^\ast}) - 1, 1 - \sigma_{\min}(S_{J_l}/\sqrt{n_l^\ast})\right\}.
	\end{split}
	\end{equation} 
	We set $\bar{A}_l = \Phi_l F_l \in \mathbb{R}^{|I_l|\times r}$, then
	$$\bar{\Sigma}_{0l} = \Phi_l F_lF_l^\top \Phi_l^\top =  \bar{A}_l\bar{A}_l^\top.$$
	On the other hand, we also recall that the true factor $A_l$ satisfies 
	\begin{equation}\label{eq:true-factor}
	A_lA_l^\top = \Sigma_{0l}^{(r)} = \Phi_l\Phi_l^\top.
	\end{equation}
	Since $\rank(\Sigma_{0l}^{(r)}) = r$ and both $A_l, \Phi_l\in \mathbb{R}^{|I_l|\times r}$, there exists an orthogonal matrix $V_l\in \mathbb{O}_r$ such that $\Phi_l = A_l V_l.$ Therefore,
	\begin{equation}\label{ineq:barA-A_l}
	\begin{split}
	& \min_{O\in \mathbb{O}_r} \|\bar{A}_l - A_lO\|_F^2 = \min_{O\in \mathbb{O}_r} \|\Phi_l F_l - \Phi_l V_l^\top O\|_F^2\\
	= & \min_{O\in \mathbb{O}_r} \|\Phi_l U_F\Sigma_F V_F^\top - \Phi_l V_l^\top O\|_F^2 \\
	\leq &  \|\Phi_l U_F\Sigma_FV_F^\top - \Phi_lV_l^\top V_lU_FV_F^\top \|_F ^2 = \|\Phi_lU_F(\Sigma_F - I_{r\times r})V_F^\top\|_F^2\\
	\leq & \|U_F(\Sigma_F - I)V_F^\top\|^2 \cdot \|\Phi_l\|_F^2 \overset{\eqref{ineq:Phi-F}}{\leq}  Cp\|\Sigma_F - I\|^2\\
	\overset{\eqref{ineq:Sigma_F-I}}{\leq} & Cp\max\left\{\sigma_{\max}(S_{J_l})/\sqrt{n_l^\ast} - 1, 1 - \sigma_{\min}(S_{J_l})/\sqrt{n_l^\ast}\right\}^2.
	\end{split}
	\end{equation}
	Let $T = \max\left\{\sigma_{\max}(S_{J_l})/\sqrt{n_l^\ast} - 1, 1 - \sigma_{\min}(S_{J_l})/\sqrt{n_l^\ast}\right\}$. Since $(\bar{\xi}_{1k}, \ldots, \bar{\xi}_{rk})$ is a sub-Gaussian vector, by random matrix theory (c.f. Theorem 5.39 in \cite{vershynin2010introduction}), 
	\begin{equation}\label{ineq:tail-bound-S}
	\begin{split}
	& \mathbb{P}\left(T\geq C\sqrt{r/n_l^\ast} + t\right)\\\
	= &  \mathbb{P}\left(\max\left\{\sigma_{\max}(S_{J_l})/\sqrt{n_l^\ast} - 1, 1 - \sigma_{\min}(S_{J_l})/\sqrt{n_l^\ast}\right\} \geq C\sqrt{r/n_l^\ast} + t\right)\\
	\leq & 1 - \mathbb{P}\Bigg(1 - C\sqrt{r/n_l^\ast} - t \leq \sigma_{\min}\left(S_{J_l}/\sqrt{n_l^\ast}\right)\leq \sigma_{\max}\left(S_{J_l}/\sqrt{n_l^\ast}\right) \\
	& \qquad \qquad \qquad \qquad \qquad \leq 1 + C\sqrt{r/n_l^\ast} + t\Bigg) \\
	\leq & C\exp(-cn_l^\ast t^2),\quad \forall t\geq 0.
	\end{split}
	\end{equation}
	Then,
	\begin{equation*}
	\begin{split}
	\mathbb{E}T^2 = & \int_{0}^\infty 2t\mathbb{P}(T\geq t)dt \leq \int_0^{C\sqrt{r/n_l^\ast}} 2t \cdot 1 \cdot dt + \int_{C\sqrt{r/n_l^\ast}}^\infty 2t\mathbb{P}\left(T \geq t\right)dt\\
	\leq & Cr/n_l^\ast + \int_{0}^\infty 2\left(t + C\sqrt{r/n_l^\ast}\right)\mathbb{P}\left(T\geq C\sqrt{r/n_l^\ast} + t\right) \\
	\leq & \int_0^\infty C\left(C\sqrt{r/n_l^\ast} + t\right)\exp(-cn_l^\ast t^2)dt\\
	= & Cr/n_l^\ast + C\sqrt{r/n_l^\ast} \cdot \sqrt{\frac{1}{n_l^\ast}} \cdot \int_{0}^\infty \exp(-ct^2)dt + \int_0^\infty \frac{C}{n_l^\ast} t\exp(-ct^2)dt\\
	\leq & Cr/n_l^\ast.
	\end{split}
	\end{equation*}
	\begin{equation}\label{ineq:bar-A-AO}
	\begin{split}
	& \mathbb{E}\min_{O\in \mathbb{O}_r} \|\bar{A}_l - A_lO\|_F^2 \\
	\overset{\eqref{ineq:barA-A_l}}{\leq} & Cp\mathbb{E}\max\left\{\sigma_{\max}(S_{J_l})/\sqrt{n_l^\ast} - 1, 1 - \sigma_{\min}(S_{J_l})/\sqrt{n_l^\ast}\right\}^2 \\
	\leq & Cp\mathbb{E}T^2 \leq Cpr/n_l^\ast.
	\end{split}
	\end{equation}
	\item[Step 3] Then we consider $\min_{O\in \mathbb{O}_r}\|\hat{A}_l - \bar{A}_lO\|_F$ in this step. We apply Lemma \ref{lm:hat_Sigma-to-hat_Sigma_0} to $\hat{A}_l\hat{A}_l^\top$ and $\bar{\Sigma}_{0l}+\sigma^2 I_{|I_l|}$. Then,
	\begin{equation}\label{ineq:Step3-1}
	\begin{split}
	\left\|\hat{A}_l\hat{A}_l^\top - \bar{\Sigma}_{0l}\right\|_F^2 \leq & C|I_l|/(|I_l|-r)\left(\|\hat{\Sigma}_l - \bar{\Sigma}_{0l} - \sigma^2I_{|I_l|}\|_F^2\right)\\
	\leq  & C\|\hat{\Sigma}_{0l} - \bar{\Sigma}_{0l} - \sigma^2I_{|I_l|}\|_F^2. 
	\end{split}
	\end{equation}
	By setting $M = \bar{\Sigma}_{0l} = \bar{A}_l\bar{A}_l^\top, \hat{M} = \hat{A}_l\hat{A}_l^\top$ in Lemma \ref{lm:factorization-lemma}, we have
	\begin{equation}\label{ineq:Step3-2}
	\begin{split}
	\min_{O\in \mathbb{O}_r} \|\hat{A}_l - \bar{A}_lO\|_F^2 \leq & \frac{\|\hat{A}_l\hat{A}_l^\top - \bar{\Sigma}_{0l} \|_F^2}{\sigma_r(\bar{A}_l)\sigma_r(\hat{A}_l)} \wedge \left(\|\hat{A}_l\|_F^2 + \|\bar{A}_l\|_F^2\right)\\
	\leq & \frac{C\|\hat{\Sigma}_{0l} - \bar{\Sigma}_{0l} - \sigma^2I_{|I_l|}\|_F^2}{\sigma_r(\bar{A}_l)\sigma_r(\hat{A}_l)} \wedge \left(\|\hat{A}_l\|_F^2 + \|\bar{A}_l\|_F^2\right).
	\end{split}
	\end{equation}
	\item[Step 4] In this step, we give an upper bound for $\mathbb{E}\|\hat{\Sigma}_{0l} - \bar{\Sigma}_{0l} - \sigma^2 I_{|I_l|}\|_F^2$. First,
	\begin{equation*}
	\begin{split}
	& \left\|\hat{\Sigma}_{0l} - \bar{\Sigma}_{0l} - \sigma^2I_{|I_l|}\right\|_F \\
	\leq & \left\|\frac{1}{n_l^\ast}\sum_{k\in J_l}\left((X_k)_{I_l} - \bar{X}_{I_l}\right)\left((X_k)_{I_l} - \bar{X}_{I_l}\right)^\top - \bar{\Sigma}_{0l} - \sigma^2I_{|I_l|}\right\|_F\\
	= & \left\|\frac{1}{n_l^\ast}\sum_{k\in J_l} (X_k)_{I_l} (X_k)_{I_l}^\top - \bar{X}_{I_l}\bar{X}_{I_l}^\top - \bar{\Sigma}_{0l} - \sigma^2I_{|I_l|}\right\|_F \\
	\leq & \Bigg\|\frac{1}{n_l^\ast} \sum_{k \in J_l}\left(\Phi_l S_{[:, k]}+ (Z_k^{(-r)})_{I_l} + (\epsilon_k)_{I_l}\right)\left(\Phi_l S_{[:, k]} + (Z_k^{(-r)})_{I_l} + (\epsilon_k)_{I_l}\right)^\top \\
	& - \frac{1}{n_l^\ast}(\Phi_l S_{[:, k]})(\Phi_l S_{[:, k]})^\top  - \sigma^2I_{|I_l|}\Bigg\|_F  + \left\|\bar{X}_{I_l}\bar{X}_{I_l}^\top\right\|_F\\
	\leq & 2\left\|\frac{1}{n_l^\ast}\sum_{k \in J_l}\Phi_l S_{[:,k]} (Z^{(-r)}_k)_{I_l}^\top\right\|_F + \frac{1}{n_l^\ast}\left\|\sum_{k\in J_l}(Z_k^{(-r)})_{I_l}(Z^{(-r)}_k)_{I_l}^\top\right\|_F \\
	& + \left\|\frac{1}{n_l^\ast}\sum_{k \in J_l} (\epsilon_k)_{I_l}(\epsilon_k)_{I_l}^\top - \sigma^2 I_{|I_l|} \right\|_F\\
	& + \frac{2}{n_l^\ast}\left\|\sum_{k\in J_l}\left(\Phi_l S_{[:,k]} + (Z_k^{(-r)})_{I_l}\right) (\epsilon_k)_{I_l}^\top\right\|_F +  \left\|\bar{X}_{I_l}\bar{X}_{I_l}^\top\right\|_F.
	\end{split}
	\end{equation*}
	We analyze each term separately.
	\begin{itemize}[leftmargin=*]
		\item 
		Since $S_{[:, k]}$ and $Z_k^{(-r)}(t)$ correspond to different scores in the Karhunen-Lo\`eve decomposition, they must be with mean zero and  uncorrelated, which implies that $\mathbb{E} \Phi_l S_{[:, k]} Z^{(-r)}_k(t) = 0$. In addition, $\left\{\Phi_l S_{[:, k]}Z_k^{(-r)}(t)\right\}$ are i.i.d. for different $k$. Thus,
		\begin{equation*}
		\begin{split}
		& \frac{1}{(n_l^\ast)^2}\mathbb{E}\left\|\sum_{k \in J_l} \Phi_l S_{[k, :]}(Z_k^{(-r)})_{I_l}^\top\right\|_F^2\\
		= & \frac{1}{(n_l^\ast)^2}\sum_{k \in J_l}\mathbb{E} \left\|\Phi_l S_{[k, :]} (Z_k^{(-r)})_{I_l}^\top\right\|_F^2 =  \frac{1}{n_l^\ast} \mathbb{E} \|\Phi_l S_{[:, 1]}(Z_1^{(-r)})_{I_l}^\top\|_F^2 \\
		= & \frac{1}{n_l^\ast} \mathbb{E}\left\{\|\Phi_l S_{[:, 1]}\|_2^2 \cdot\|(Z^{(-r)}_1)_{I_l}\|_2^2\right\} \leq \frac{1}{n_l^\ast} \left(\mathbb{E}\|\Phi_l S_{[:, 1]}\|_2^4 \cdot \mathbb{E} \|(Z_1^{(-r)})_{I_l}\|_2^4\right)^{1/2}.\\
		\end{split}
		\end{equation*}
		Here,
		\begin{equation*}
		\begin{split}
		& \mathbb{E}\|(Z_1^{(-r)})_{I_l}\|_2^4 \\
		= & \mathbb{E}\left(\sum_{i \in I_l} Z_1^{(-r)}(T(I_l(i)))^2 \right)^2 \overset{\text{Cauchy-Schwarz}}{\leq} |I_l| \sum_{i\in I_l} Z_1^{(-r)}(T(I_l(i)))^4 \\
		\leq & |I_l|^2 \cdot \sup_t \mathbb{E}(Z_1^{(-r)}(t))^4  \leq Cp^2r/(n^\ast \gamma),
		\end{split}
		\end{equation*}
		where the last inequality is due to the assumption of this proposition.
		\begin{equation*}
		\begin{split}
		& \mathbb{E} \|\Phi_l S_{[:, 1]}\|_2^4 \\
		= & \mathbb{E} \|X_{I_l} - (Z^{(-r)}_k)_{I_l} - (\epsilon_k)_{I_l}\|_2^4\\
		\leq & C\left(\mathbb{E} \|X_{I_l}\|_2^4 + \|(Z^{(-r)}_k)_{I_l}\|_2^4 + (\epsilon_k)_{I_l}\|_2^4\right)\\
		\leq & C|I_l|\left(\sum_{i \in I_l}\mathbb{E}|X(T(I_l(i)))|^4 + \sum_{i \in I_l}\mathbb{E}|Z^{(-r)}(T(I_l(i)))|^4 + \sum_{i \in I_l}\mathbb{E}|\epsilon(T(I_l(i)))|^4\right)\\
		\leq & C|I_l|^2 \left(\sup_t \mathbb{E}X(t)^4 + \sup_t \mathbb{E}Z^{(-r)}(t)^4 +  \mathbb{E}\epsilon^4\right) \leq Cp^2.
		\end{split}
		\end{equation*}
		Provided that $n^\ast \geq C\gamma^2\geq Cr\gamma$, we have
		\begin{equation*}
		\frac{1}{(n_l^\ast)^2}\mathbb{E}\left\|\sum_{k \in J_l} \Phi_l S_{[k, :]}(Z_k^{(-r)})_{I_l}^\top\right\|_F^2\leq \frac{C}{n_l^\ast}\left(p^2r/(n_l^\ast\gamma) \cdot p^2\right)^{1/2} \leq \frac{Cp^2r}{\gamma n_l^\ast}.
		\end{equation*}
		\item With the assumption that $(\mathbb{E}\epsilon^4)^{1/2} \leq Cr/\gamma$, we have
		\begin{equation}\label{ineq:Eepsilon^4}
		\begin{split}
		\mathbb{E}\|(\epsilon_k)_{I_l}\|_2^4 = & \mathbb{E}\left(\sum_{i\in I_l} \epsilon_k(T(I_l(i)))^2 \right)^2 \overset{\text{Cauchy-Schwarz}}{\leq} |I_l|\cdot \sum_{i\in I_l} \mathbb{E}\epsilon_k(T(I_l(i)))^4 \\
		\leq & Cp^2 (r/\gamma)^2.
		\end{split}
		\end{equation}
		\begin{equation*}
		\sigma^4 = \left(\mathbb{E} \epsilon^2\right)^2 \leq \mathbb{E} \epsilon^4 \leq C(r/\gamma)^2.
		\end{equation*}
		Given $\mathbb{E}(\epsilon_k)_{I_l}(\epsilon_k)_{I_l}^\top - \sigma^2I_{|I_l|} = 0$, we have
		\begin{equation*}
		\begin{split}
		& \mathbb{E}\left\|\frac{1}{n_l^\ast}\sum_{k \in J_l} (\epsilon_k)_{I_l}(\epsilon_k)_{I_l}^\top - \sigma^2I_{|I_l|}\right\|_F^2 = \frac{1}{(n_l^\ast)^2} \sum_{k \in J_l} \mathbb{E}\left\|(\epsilon_k)_{I_l}(\epsilon_k)_{I_l}^\top - \sigma^2 I_{|I_l|}\right\|_F^2\\
		= & \frac{1}{n_l^\ast} \mathbb{E}\left\|\epsilon_{I_l}\epsilon_{I_l}^\top - \sigma^2 I_{|I_l|}\right\|_F^2 \leq \frac{2}{n_l^\ast} \left(\mathbb{E}\left\|\epsilon_{I_l}\epsilon_{I_l}^\top \right\|_F^2 + \left\|\sigma^2 I_{|I_l|}\right\|_F^2\right)\\ 
		= &  \frac{2}{n_l^\ast}\left(\mathbb{E}\|\epsilon_{I_l}\|_2^4 + \sigma^4 p\right) \leq \frac{2}{n_l^\ast} \left(|I_l|\cdot \sum_{i\in I_l} \mathbb{E} \epsilon_k(T(I_l(i)))^4 + \sigma^4 p \right) \leq \frac{Cp^2r}{n_l^\ast \gamma}.
		\end{split}
		\end{equation*}
		\item With the assumption that $\sup_{t\in \mathcal{T}}\mathbb{E}X(t)^4 \leq C$, and $X_k(t) = Z_k(t)  + \epsilon_{k}(t)$, we have
		\begin{equation}\label{ineq:EX^4}
		\begin{split}
		\mathbb{E}\|(Z_k)_{I_l}\|_2^4 \leq & C\mathbb{E}\|(X_k)_{I_l}\|_2^4 + C\mathbb{E}\|(\epsilon_k)_{I_l}\|_2^4 \\
		\leq & C\mathbb{E}\left(\sum_{i\in I_l} X_k(T(I_l(i)))^2 \right)^2 + C\mathbb{E}\left(\sum_{i\in I_l} \epsilon_k(T(I_l(i)))^2 \right)^2\\
		\overset{\text{Cauchy-Schwarz}}{\leq} & C|I_l|\cdot \sum_{i\in I_l} \mathbb{E}X_k(T(I_l(i)))^4 + C|I_l|\cdot \sum_{i\in I_l} \mathbb{E}\epsilon_k(T(I_l(i)))^4\\
		\leq & C|I_l|^2 \leq Cp^2.
		\end{split}
		\end{equation}
		Given $\mathbb{E} \epsilon_k =0$, $\epsilon_k$ and $(S_k, Z^{(-r)}_k(t))$ are uncorrelated, we have\\ $\mathbb{E}\left(\Phi_l S_{[:, k]} + (Z_k^{(-r)})_{I_l}\right) (\epsilon_k)_{I_l}^\top = 0$ and
		\begin{equation*}
		\begin{split}
		& \mathbb{E}\frac{1}{(n_l^\ast)^2}\left\|\sum_{k \in J_l}\left(\Phi_l S_{[:, k]} + (Z_k^{(-r)})_{I_l}\right) (\epsilon_k)_{I_l}^\top\right\|_F^2 \\
		= & \mathbb{E} \frac{1}{n_l^\ast} \mathbb{E} \left\|(Z_k)_{I_k} \cdot (\epsilon_k)_{I_k}^\top\right\|_F^2\\
		= & \mathbb{E}\frac{1}{n_l^\ast} \|(Z_k)_{I_k}\|_2^2 \cdot \|(\epsilon_k)_{I_k}\|_2^2 \leq \frac{1}{n_l^\ast} \left(\mathbb{E} \|(Z_k)_{I_k}\|_2^4 \cdot \mathbb{E}\|(\epsilon_k)_{I_k}\|_2^4\right)^{1/2}\\ \overset{\eqref{ineq:Eepsilon^4}\eqref{ineq:EX^4}}{\leq} & \frac{Cp^2r}{n_l^\ast\gamma}.
		\end{split}
		\end{equation*}
		\item Given $\mathbb{E}X_k(t) = 0$ and $X_1(t),\ldots, X_n(t)$ are independent, 
		\begin{equation*}
		\begin{split}
		& \mathbb{E} \left\|\bar{X}_{I_l}\bar{X}_{I_l}^\top\right\|_F^2 = \frac{1}{(n_l^\ast)^4}\mathbb{E}\left\|\sum_{k\in J_l} (X_k)_{I_l} \right\|_2^4 \leq \frac{C}{(n_l^\ast)^2} \mathbb{E} \|(X_k)_{I_l}\|_2^4\\
		\leq & \frac{C}{(n_l^\ast)^2} \cdot |I_l|\cdot  \sum_{i\in I_l} \mathbb{E} X_k(T(I_l(i)))^4 \leq \frac{C|I_l|^2}{(n_l^\ast)^2} \leq \frac{Cp^2}{(n_l^\ast)^2} \leq \frac{Cp^2}{n_l^\ast} \cdot \frac{r}{\gamma}.
		\end{split}
		\end{equation*}
	\end{itemize}
	In summary,
	\begin{equation}\label{ineq:hat_Sigma-Sigma-sigma}
	\begin{split}
	\mathbb{E}\left\|\hat{\Sigma}_{0l} - \bar{\Sigma}_{0l} - \sigma^2I_{|I_l|} \right\|_F^2 \leq & \frac{Cp^2r}{\gamma n_l^\ast}.
	\end{split}
	\end{equation}
	\item[Step 4] In this step, we further introduce the following ``good" event,
	\begin{equation}\label{eq:Bast-good-event}
	B_\ast = \left\{\sigma_r^2(\hat{A}_l) \geq \sigma_r^2(A_l)/4, \sigma_r^2(\bar{A}_l) \geq \sigma_r^2(A_l)/2, \forall 1\leq l\leq l_{\max}\right\}.
	\end{equation}
	Then we develop the upper bound under this good event to finalize the proof. First, we aim to show $B^\ast$ happens with high chance. By \eqref{ineq:Step3-1}, we have 
	\begin{equation*}
	\begin{split}
	& \left\|\hat{A}_l\hat{A}_l^\top - \bar{A}_l\bar{A}_l^\top\right\|_F \leq C\left\|\hat{\Sigma}_{0l} - \bar{\Sigma}_{0l} - \sigma^2 I_{|I_l|}\right\|_F\\
	\Rightarrow \quad & \sigma_r^2(\hat{A}_l) \geq \sigma_r^2(\bar{A}_l) - C\|\hat{\Sigma}_{0l} - \bar{\Sigma}_{0l} - \sigma^2 I_{|I_l|}\|_F.
	\end{split}
	\end{equation*}
	By definition, 
	$$\sigma_r^2(\bar{A}_l) = \sigma_r\left(\frac{1}{n_l^\ast}\Phi_l S_{J_l}S_{J_l}^\top \Phi_l^\top\right) \geq \sigma_{\min}^2(\Phi_l)\sigma_r^2(S_{J_l}/\sqrt{n_l}). $$
	In addition, 
	$$\sigma_r^2(A_l) \overset{\eqref{eq:true-factor}}{=} \sigma_{\min}^2(\Phi_l) \overset{\eqref{eq:sigma_r-Phi}}{\geq} cp/\gamma.$$
	Thus, $B_\ast$ holds if the following two conditions hold for some small constant $c>0$:
	\begin{equation}\label{eq:condition}
	\forall l, \quad \sigma_r^2(S_{J_l}/\sqrt{n_l}) \geq 1/2, \quad \text{and}\quad \|\hat{\Sigma}_{0l} - \bar{\Sigma}_{0l}-\sigma^2I_{|I_l|}\|_F \leq cp/\gamma.
	\end{equation}
	By Markov's inequality and the sub-Gaussian random matrix tail bound \eqref{ineq:tail-bound-S},
	\begin{equation}\label{ineq:PB-ast}
	\begin{split}
	& \mathbb{P}(B_\ast \text{ holds}) \geq \mathbb{P}(\eqref{eq:condition} \text{ holds}) \\
	\geq & 1 - \mathbb{P}\left(\exists l, \sigma_r^2(S_{J_l}/\sqrt{n_l}) < 1/2\right) + \mathbb{P}\left(\exists l, C\|\hat{\Sigma}_{0l} - \bar{\Sigma}_{0l}-\sigma^2I_{|I_l|}\|_F \geq cp/\gamma\right) \\
	\overset{\eqref{ineq:tail-bound-S}}{\geq} & 1 - l_{\max} \exp(-cn_l^\ast) - l_{\max} \frac{\mathbb{E}\|\hat{\Sigma}_{0l}-\bar{\Sigma}_{0l}-\sigma^2I_{|I_l|\|_F^2}\|_F^2}{c(p/\gamma)^2}\\ \overset{\eqref{ineq:hat_Sigma-Sigma-sigma}}{\geq} & 1 - C\exp(-cn^\ast) - Cr\gamma/n^\ast\\
	\geq & 1 - Cr\gamma/n^\ast.
	\end{split}
	\end{equation}
	When $B_\ast$ holds, we must have
	\begin{equation*}
	\sigma_r(A_l), \sigma_r(\bar{A}_l), \sigma_r(\hat{A}_l)\geq c\sqrt{p/\gamma}.
	\end{equation*}
	By combining \eqref{ineq:Step3-2}, \eqref{ineq:hat_Sigma-Sigma-sigma}, and the previous inequality, we have for all $1\leq l \leq l_{\max}$,
	\begin{equation}\label{ineq:hat-A-l-bar-A}
	\begin{split}
	\min_{O\in \mathbb{O}_r} \mathbb{E}\|\hat{A}_l - \bar{A}_lO\|_F^2 1_{\{B_\ast \text{ holds}\}} \leq & \frac{C\mathbb{E}\|\hat{\Sigma}_{0l} - \bar{\Sigma}_{0l} - \sigma^2I_{|I_l|}\|_F^2}{\sigma_r(\bar{A}_l)\sigma_r(\hat{A}_l)}\cdot 1_{\{B_\ast \text{ holds}\}} \\
	\leq & \frac{C\mathbb{E}\|\hat{\Sigma}_{0l} - \bar{\Sigma}_{0l} - \sigma^2I_{|I_l|}\|_F^2}{p/\gamma} \leq \frac{Cpr}{n_l^\ast}.
	\end{split}
	\end{equation}
\end{enumerate}
Finally, \eqref{ineq:triangle}, \eqref{ineq:bar-A-AO}, and \eqref{ineq:hat-A-l-bar-A} conclude the statement of this lemma. \quad $\square$

Now we consider the proof of Proposition \ref{pr:complete}. Similarly to the proof of Theorem \ref{th:upper_bound_func}, we develop an upper bound on the probability of the ``bad case," i.e., $B_\ast$ does not hold. To this end, we define $w\in \mathbb{R}^{p}, w_i = |\{l: i\in I_l\}|^{-1}$ as the weight in Equation \eqref{eq:tilde-A}. Then,
\begin{equation*}
\begin{split}
\|\hat{\Sigma}_0\|_F \leq &  \|\tilde{A}\tilde{A}^\top\|_F = \left\|\diag(w)\left(\sum_{l=1}^{l_{\max}}\hat{A}_l^\ast\right)\left(\sum_{l=1}^{l_{\max}}\hat{A}_l^\ast\right)^\top \diag(w)\right\|_F \\
\leq & \left\|\left(\sum_{l=1}^{l_{\max}}\hat{A}_l^\ast\right)\left(\sum_{l=1}^{l_{\max}}\hat{A}_l^\ast\right)^\top \right\|_F  \leq l_{\max} \cdot \sum_{l=1}^{l_{\max}} \|\hat{A}_l^\ast (\hat{A}_l^\ast)^\top\|_F \\
= & l_{\max}\sum_{l=1}^{l_{\max}}\|\hat{A}_l\hat{A}_l^\top\|_F \overset{\eqref{eq:hat-A_l}}{\leq} l_{\max} \sum_{l=1}^{l_{\max}} \|\hat{\Sigma}_l\|_F
\end{split}
\end{equation*}
Then,
\begin{equation*}
\begin{split}
\mathbb{E}\|\hat{\Sigma}_{l}\|_F^2 = & \mathbb{E}\left\|\frac{1}{n_l^\ast}\sum_{k\in J_l}((X_k)_{I_l} - \bar{X}_{I_l})((X_k)_{I_l} - \bar{X}_{I_l})^\top \right\|_F^2\\
= & \mathbb{E} \left\|\frac{1}{n_l^\ast}\sum_{k \in J_l}(X_k)_{I_l}(X_k)_{I_l}^\top - \bar{X}_{I_l} \bar{X}_{I_l}\right\|_F^2\\
\overset{\text{\text{Cauchy-Schwarz}}}{\leq} & \frac{C}{n_l^\ast} \sum_{k \in J_l} \mathbb{E} \|(X_k)_{I_l} (X_k)_{I_l}^\top\|_F^2 + C\mathbb{E}\|\bar{X}_{I_l}\bar{X}_{I_l}\|_F^2 \\
= & \frac{C}{n_l^\ast} \sum_{k \in J_l} \mathbb{E} \|(X_k)_{I_l}\|_2^4 + C\mathbb{E}\|\bar{X}_{I_l}\|_2^4 \overset{\eqref{ineq:EX^4}}{\leq} Cp^2. 
\end{split}
\end{equation*}
\begin{equation*}
\mathbb{E}\|\hat{\Sigma}_0 - \Sigma_0\|_F^2 \leq C\mathbb{E}\|\hat{\Sigma}_0\|_F^2 +  C\|\Sigma_0\|_F^2 \leq Cp^2.
\end{equation*}
By Cauchy-Schwarz inequality,
\begin{equation*}
\begin{split}
\mathbb{E}\|\hat{\Sigma}_0 - \Sigma_0\|_F 1_{B_\ast^c} \leq & \left(\mathbb{E}\|\hat{\Sigma}_0 - \Sigma_0\|_F^2 \cdot \mathbb{E} 1_{B_\ast^c}^2\right)^{1/2}\\
\overset{\eqref{ineq:PB-ast}}{\leq} & \left(Cp^2 \cdot \gamma r/n^\ast\right)^{1/2}.
\end{split}
\end{equation*}
Similarly to Steps 3 - 5 and based on Lemma \ref{lm:factor-estimation}, one can develop the upper bound for $\|\hat{\Sigma}_0 - \Sigma\|_F$ on the ``good event,"
\begin{equation*}
\mathbb{E}\|\hat{\Sigma}_0 - \Sigma\|_F \cdot 1_{\{B_\ast \text{ holds}\}}\leq C\sqrt{p^2r\gamma/n^\ast}.
\end{equation*}
Thus,
\begin{equation*}
\mathbb{E}\|\hat{\Sigma}_0 - \Sigma\|_F = \mathbb{E}\|\hat{\Sigma}_0 - \Sigma\|_F1_{\{B_\ast \text{ holds}\}} + \mathbb{E}\|\hat{\Sigma}_0 - \Sigma\|_F1_{\{B_\ast^c \text{ holds}\}} \leq C\sqrt{p^2r\gamma/n^\ast}.
\end{equation*}
Finally, since $\Sigma_0$ is a $p$-by-$p$ linear interpolation for $G$, we finally have
\begin{equation*}
\left\|\hat{G} - G\right\|_{HS} \leq \frac{1}{p} \|\hat{\Sigma}_0 - \Sigma_0\|_F + O(p^{-1}) = O(\sqrt{\gamma r/n^\ast}+p^{-1}),
\end{equation*}
which has finished the proof of Proposition \ref{pr:complete}. \quad $\square$


\section{Technical Lemmas}

We collect all technical tools that were used in the main context of this paper in this section. We first provide the proof for Lemma \ref{lm:rotations-Wahba}, which provides an error bound for Wahba's problem \citep{wahba1965least}. 

{\bf\noindent Proof of Lemma \ref{lm:rotations-Wahba}.} Based on our assumption,
\begin{equation*}
\begin{split}
& \left\|A_2\hat{O} - A_1\right\|_F \leq \left\| A_2O_2^\top O_1 - A_1 \right\|_F = \left\|A_2O_2^\top  - A_1 O_1^\top \right\|_F\\
\leq & \left\|A_2O_2^\top  - A\right\|_F + \left\|A_1O_1^\top - A \right\|_F = \|A_2 - AO_2\|_F + \|A_1 - AO_1\|_F \\
\leq & a_1+a_2.
\end{split}
\end{equation*}
On the other hand,
\begin{equation*}
\begin{split}
& \left\|A_2\hat{O} - A_1\right\|_F \\
\geq &  -\|A_1 - AO_1\|_F - \|A_2\hat{O} - AO_2 \hat{O}\|_F + \|AO_1 - AO_2\hat{O}\|_F\\
\geq & -a_1 - a_2 + \sigma_{\min}(A) \|O_1 - O_2\hat{O}\|_F\\
\geq & -a_1 -a_2 + \lambda \|\hat{O} - O_2^\top O_1\|_F.
\end{split}
\end{equation*}
Therefore,
\begin{equation*}
\left\|\hat{O} - O_2^\top O_1 \right\|_F\leq \frac{2(a_1+a_2)}{\lambda}.
\end{equation*}
Finally, 
\begin{equation*}
\begin{split}
\left\|O_2 \hat{O} - I\right\|_F \leq \left\|O_2 \hat{O} - O_1 \right\|_F + \left\|O_1 - I \right\|_F \leq \|O_1 - I\|_F + \frac{2(a_1+a_2)}{\lambda}.
\end{split}
\end{equation*}
\quad $\square$

The following lemma characterizes the least and largest singular value of semi-positive symmetric definite matrix factorization.
\begin{Lemma}\label{lm:algebra-least-singular-value}
	Suppose a positive semidefinite matrix $A \in \mathbb{R}^{p\times p}$ can be decomposed as $A = HDH^\top$. Here $D\in \mathbb{R}^{r\times r}$ is a non-negative diagonal matrix and $H\in \mathbb{R}^{p\times r}$ is a general matrix that is not necessarily orthogonal. Then
	$$\left(\max_i D_{ii}\right)\sigma_r^2(H) \geq \sigma_r(A)\geq \left(\min_i D_{ii}\right)\sigma_r^2(H),$$
	$$\left(\max_{i}D_{ii}\right)\|H\|^2 \geq \|A\| \geq \left(\min_{i}D_{ii}\right)\|H\|^2.$$
\end{Lemma}
{\bf\noindent Proof of Lemma \ref{lm:algebra-least-singular-value}.} Suppose the singular value decomposition of $H$ is $H = U_H D_H V_H^\top$, where $U_H\in \mathbb{O}_{p, r}$, $D_H\in \mathbb{R}^{r\times r}$ is diagonal with non-increasing non-negative entries, $V_H\in \mathbb{O}_{p, r}$. Then, 
\begin{equation*}
\begin{split}
\|A\| =&  \max_{\|u\|_2\leq 1} u^\top Au = \max_{\|u\|_2\leq 1} u^\top HDH^\top u \geq \left(U_{H, [:, 1]}^\top H\right) D \left(H^\top U_{H, [:, 1]}\right) \\
\geq & \sigma_r(D) \cdot \left\|H^\top U_{H, [:, 1]}\right\|_2^2 = \min_{1\leq i \leq r} D_{ii} \cdot \sigma_1^2(H),
\end{split}
\end{equation*}
\begin{equation*}
\begin{split}
\|A\| = & \max_{\|u\|_2\leq 1}u^\top A u = \max_{\|u\|_2\leq 1}u^\top HDH^\top u \\
\leq & \max_{\|u\|_2\leq 1}\|u\|_2 \cdot\|H\|\cdot\|D\|\cdot\|H^\top\| \cdot\|u\|_2 = \left(\max_{i}D_{ii}\right) \|H\|^2.
\end{split}
\end{equation*}
On the other hand, without loss of generality we assume $D_{rr} = \min_i D_{ii}$, then
\begin{equation*}
\begin{split}
\sigma_r(A) = & \sigma_r\left(U_H\left(D_HV_H^\top DV_HD_H\right)U_H^\top\right) = \sigma_r\left(D_HV_H^\top DV_HD_H\right)\\
= & \min_{u\in \mathbb{R}^r: \|u\|_2= 1} u^\top D_HV_H^\top DV_HD_H u \leq e_r^\top D_H V_H^\top D V_H D_H e_r \\
\leq & \|D\|\cdot \|e_r^\top D_H V_H^\top\|_2^2 = \left(\max_i D_{ii}\right)\sigma_r^2(H),
\end{split}
\end{equation*}
\begin{equation*}
\begin{split}
\sigma_r(A) = & \sigma_r\left(U_H\left(D_HV_H^\top DV_HD_H\right)U_H^\top\right) = \sigma_r\left(D_HV_H^\top DV_HD_H\right)\\
\geq &  \sigma_{\min}^2(D_HV_H^\top)\sigma_{\min}(D) = \left(\min_{i}D_{ii}\right)\sigma_r^2(H).
\end{split}
\end{equation*}
These have finished the proof for this lemma. \quad $\square$

\begin{Lemma}\label{lm:hat_Sigma-to-hat_Sigma_0}
	Suppose $\Sigma = \Sigma_0 + \sigma^2I\in \mathbb{R}^{b\times b}$. Here, $\Sigma_0$ is positive semi-definite, $\Sigma_0 = \Sigma_0^{(r)} + \Sigma_0^{(-r)}$, $\Sigma_0^{(r)}$ is a rank-$r$ matrix. Suppose $\hat{\Sigma}$ is another rank-$r$ symmetric matrix  satisfying $\|\hat{\Sigma} - \Sigma\|_F\leq \lambda$. Suppose $\hat{U}\hat{D}\hat{U}^\top$ is the eigenvalue decomposition and
	\begin{equation}
	\hat{\Sigma}_0 = \sum_{i=1}^r \hat{U}_{[:, i]} \left\{(\hat{D}_{ii} - \hat{\sigma}^2) \vee 0\right\} (\hat{U}_{[:, i]})^\top,  \text{ where } \hat{\sigma}^2 = \left(\frac{1}{b-r} \sum_{i=r+1}^b\hat{D}_{ii}\right)\vee 0, 
	\end{equation}
	then the following inequality holds,
	\begin{equation}
	\left\|\hat{\Sigma}_0 - \Sigma_0\right\|_F \leq C\sqrt{b/(b-r)}\left(\lambda + \|\Sigma_0^{(-r)}\|_F\right).
	\end{equation}
	for uniform constant $C>0$.
\end{Lemma}

{\noindent\bf Proof of Lemma \ref{lm:hat_Sigma-to-hat_Sigma_0}.} 
Since $\hat{\Sigma}=\sum_{i=1}^b\hat{U}_{[:, i]}\hat{D}_{ii}\hat{U}_{[:, i]}^\top$ is the eigenvalue decomposition of $\hat{\Sigma}$, we also have the following eigenvalue decomposition for $\hat{\Sigma} - \sigma^2I_{b\times b}$,
\begin{equation*}
\hat{\Sigma} - \sigma^2 I_{b\times b} = \Sigma^{(r)} + \Sigma_0^{(-r)} + (\hat{\Sigma} - \Sigma) = \sum_{i=1}^{b}\hat{U}_{[:, i]}(\hat{D}_{ii}-\sigma^2)\hat{U}_{[:, i]}^\top.
\end{equation*}
Additionally, since $\Sigma_0^{(r)}$ is positive semi-definite, we can write down the eigenvalue decomposition $\Sigma_0^{(r)} = \sum_{i=1}^r U_{[:, i]}D_{ii}U_{[:, i]}^\top$, where $U\in \mathbb{O}_{b, r}$, $D\in \mathbb{R}^{r\times r}$ is non-negative diagonal. By Lemma \ref{lm:A-B}, 
\begin{equation}\label{ineq:hat-D-ii-sigma^2}
\begin{split}
& \|\{\hat{D}_{ii} - \sigma^2\}_{i=r+1}^b\|_2 = \left(\sum_{i=r+1}^b \left(\hat{D}_{ii}-\sigma^2\right)^2\right)^{1/2} \\
\leq & \left(\sum_{i=1}^r (\hat{D}_{ii} - \sigma^2 - D_{ii})^2 + \sum_{i=r+1}^b (\hat{D}_{ii}-\sigma^2)^2\right)^{1/2}\\
\leq & \|\hat{\Sigma}-\sigma^2I_{b\times b}-\Sigma_0^{(r)}\|_F = \|\Sigma_0^{(-r)} + \hat\Sigma - \Sigma \|_F \leq \lambda + \|\Sigma_0^{(-r)}\|_F.
\end{split}
\end{equation}
Then
\begin{equation}\label{ineq:lm-hat-sigma0-1}
\begin{split}
\left|\hat\sigma^2 - \sigma^2 \right| = & \left|\left(\frac{1}{b-r}\sum_{i=r+1}^b \hat{D}_{ii}\right)\vee 0 - \sigma^2 \right| \leq \left|\left(\frac{1}{b-r}\sum_{i=r+1}^b \hat{D}_{ii}\right) - \sigma^2 \right|\\
\leq & \frac{1}{b-r} \sum_{i=r+1}^b\left|\hat{D}_{ii} - \sigma^2\right|\leq \frac{1}{\sqrt{b-r}} \left(\sum_{i=r+1}^b(\hat{D}_{ii}-\sigma^2)^2\right)^{1/2} \\
\leq & \frac{1}{\sqrt{b-r}}\left(\lambda + \|\Sigma_0^{(-r)}\|_F\right).
\end{split}
\end{equation}
Thus 
\begin{equation}\label{ineq:lm-hat-sigma0-2}
\begin{split}
& \left\|(\hat{\Sigma} - \hat{\sigma}^2I_{b\times b}) - \Sigma_0 \right\|_F \leq \left\|\hat{\Sigma} - \Sigma\right\|_F + \|\hat{\sigma}^2I_{b\times b} - \sigma^2I_{b\times b}\|_F \\
\leq & \sqrt{b/(b-r)}\left(\lambda + \|\Sigma_0^{(-r)}\|_F\right) + \lambda.
\end{split}
\end{equation}
On the other hand, note that $\hat{\Sigma}_0 - \hat{\sigma}^2I_{b\times b} = \sum_{i=1}^b \hat{U}_{[:, i]} (\hat{\Sigma}_{ii} - \hat{\sigma}^2) \hat{U}_{[:, i]}^\top$ and $\hat{U}_{[:, 1]}, \ldots, \hat{U}_{[:, b]}$ are orthonormal, the following inequality holds,
\begin{equation}
\begin{split}
\left\|\hat{\Sigma}_0 - \Sigma_0\right\|_F \leq & \left\|\hat{\Sigma}_0 - (\hat{\Sigma}-\hat{\sigma}^2I_{b\times b})\right\|_F + \left\|\hat{\Sigma} - \hat{\sigma}^2I_{b\times b} - \Sigma_0\right\|_F\\
\overset{\eqref{ineq:lm-hat-sigma0-2}}{\leq} & \left\|\sum_{i=1}^r \hat{U}_{[: ,i]}\left\{(\hat{D}_{ii} - \hat{\sigma}^2)\vee 0\right\} \hat{U}_{[: ,i]}^\top - \sum_{i=r+1}^b \hat{U}_{[:. i]}(\hat{D}_{ii} - \hat{\sigma}^2)\hat{U}_{[:. i]}^\top \right\|_F \\
& + \sqrt{b/(b-r)}\left(\lambda + \|\Sigma_0^{(-r)}\|_F\right).\\
\end{split}
\end{equation}
In particular,
\begin{equation}
\begin{split}
& \left\|\sum_{i=1}^r \hat{U}_{[: ,i]}\left\{(\hat{D}_{ii} - \hat{\sigma}^2)\vee 0\right\} \hat{U}_{[: ,i]}^\top - \sum_{i=r+1}^b \hat{U}_{[:. i]}(\hat{D}_{ii} - \hat{\sigma}^2)\hat{U}_{[:. i]}^\top \right\|_F^2 \\
= & \sum_{i=1}^r\left\{\{(\hat{D}_{ii} - \hat{\sigma}^2) \vee 0\} - (\hat{D}_{ii} - \hat{\sigma}^2)\right\}^2 +  \sum_{i = r+1}^b\left|\hat{D}_{ii} - \hat{\sigma}^2\right|^2.
\end{split}
\end{equation}
Here, we note that the $i$-th eigenvalue of $\Sigma$ satisfies $\lambda_i(\Sigma) = D_{ii} + \sigma^2, D_{ii}\geq 0$ for $1\leq i \leq r$, so
\begin{equation*}
\begin{split}
& \sum_{i =1}^r\left\{\{(\hat{D}_{ii} - \hat{\sigma}^2) \vee 0\} - (\hat{D}_{ii} - \hat{\sigma}^2)\right\}^2 = \sum_{1\leq i\leq r} \left\{\left(\hat{\sigma}^2 - \hat{D}_{ii}\right)_+\right\}^2\\
\leq & 3\sum_{1\leq i\leq r}\left\{\left(\hat{\sigma}^2 - \sigma^2\right)_+\right\}^2 + 3\sum_{1\leq i\leq r}\left\{\left(\sigma^2 - (D_{ii} + \sigma^2)\right)_+\right\}^2\\
& + 3\sum_{1\leq i\leq r} \left\{\left(\lambda_i(\Sigma) - \hat{D}_{ii}\right)_+\right\}^2\\
\overset{\eqref{ineq:lm-hat-sigma0-1}}{\leq} & \frac{3r}{b-r}\left(\lambda + \|\Sigma_0^{(-r)}\|_F\right)^2 + 0 + 3\sum_{i=1}^b\left\{\lambda_i(\Sigma) - \hat{D}_{ii}\right\}^2 \\
\overset{\text{Lemma \ref{lm:A-B}}}{\leq} & \frac{3b}{b-r}\left(\lambda+\|\Sigma_0^{(-r)}\|_F\right)^2 + 3\|\hat{\Sigma} - \Sigma\|_F^2 \leq \frac{3b}{b-r}\left(\lambda + \|\Sigma_0^{(-r)}\|_F\right)^2 + 3\lambda^2;
\end{split}
\end{equation*}
\begin{equation*}
\begin{split}
& \sum_{r+1\leq i \leq b}\left|\hat{D}_{ii} - \hat{\sigma}^2\right|^2 \leq \sum_{r+1\leq i \leq b} \left\{2\left|\hat{D}_{ii} - \sigma^2\right|^2 + 2\left|\sigma^2 - \hat{\sigma}^2\right|^2\right\}\\
\overset{\eqref{ineq:hat-D-ii-sigma^2}\eqref{ineq:lm-hat-sigma0-1}}{\leq} & 2\left(\lambda+ \|\Sigma_0^{(-r)}\|_F\right)^2 + \frac{2b}{b-r}\left(\lambda + \|\Sigma_0^{(-r)}\|_F\right)^2.
\end{split}
\end{equation*}
In summary, we have
$$\left\|\hat{\Sigma}_0 - \Sigma_0\right\|_F \leq C\sqrt{b/(b-r)}\left(\lambda+\|\Sigma_0^{(-r)}\|_F\right).$$
for some uniform constant $C>0$.
\quad $\square$

\begin{Lemma}\label{lm:A-B}
	Suppose $A, B\in \mathbb{R}^{d\times d}$ are two symmetric matrices. $\lambda_j(A)$ and $\lambda_j(B)$ represent the $j$-th eigenvalues of $A$ and $B$, respectively. Then
	\begin{equation}
	\|A - B\|_F^2 \geq \sum_{j=1}^d \left(\lambda_j(A) - \lambda_j(B)\right)^2.
	\end{equation}
\end{Lemma}

{\bf\noindent Proof of Lemma \ref{lm:A-B}.} Since 
\begin{equation*}
\begin{split}
\|A-B\|_F^2 = & \tr\left((A-B)^\top(A-B)\right) \\
= & \|A\|_F^2 + \|B\|_F^2 - 2\tr(A^\top B) = \sum_{j=1}^d \lambda_j^2(A) + \sum_{j=1}^2(B) - 2\tr(A^\top B),
\end{split}
\end{equation*}
we only need to show
\begin{equation}\label{ineq:to-show-1}
\tr(A^\top B) \leq \sum_{j=1}^d \lambda_j(A) \lambda_j(B).
\end{equation}
Suppose the eigenvalue decomposition of $B$ is $B = UDU^\top$, with $D = \diag(\lambda_1(B), \ldots, \lambda_d(B))$. Let $U_{\{j\}} = U_{[:, 1:j]}$, then
\begin{equation*}
B = \sum_{j=1}^d U_{[:, j]}U_{[:, j]}^\top \cdot \lambda_j(B) = \sum_{j=1}^d U_{\{j\}}U_{\{j\}}^\top \cdot \left(\lambda_j(B) - \lambda_{j+1}(B)\right),
\end{equation*}
thus,
\begin{equation*}
\begin{split}
\tr(A^\top B) = & \sum_{j=1}^d \tr\left(A^\top U_{\{j\}}U_{\{j\}}^\top\right) \cdot \left(\lambda_j(B) - \lambda_{j+1}(B)\right)\\
\overset{\text{Lemma \ref{lm:simple}}}{\leq} & \sum_{j=1}^d \tr\left(\sum_{i=1}^j \lambda_i(A)\right) \left(\lambda_j(B) - \lambda_{j+1}(B)\right)\\
= & \sum_{j=1}^d \lambda_j(A)\lambda_j(B),
\end{split}
\end{equation*}
which has finished the proof of this lemma.
\quad $\square$

\begin{Lemma}\label{lm:simple}
	Suppose $A\in \mathbb{R}^{d\times d}$ is symmetric, $U_{\{j\}}\in \mathbb{O}_{d, j}$, then
	\begin{equation*}
	\tr\left(A^\top U_{\{j\}}U_{\{j\}}^\top\right) \leq \sum_{i=1}^j \lambda_i(A)
	\end{equation*}
\end{Lemma}
{\bf \noindent Proof of Lemma \ref{lm:simple}.} Without loss of generality we can assume $A = \diag(\lambda_1(A), \ldots, \lambda_d(A))$. Since $U_{\{j\}}\in \mathbb{O}_{d, j}$, we have
$$0\leq (U_{\{j\}})_{ii} \leq 1,\quad  \sum_{i=1}^d (U_{\{j\}})_{ii} = j,$$
then by rearrangement inequality,
\begin{equation*}
\begin{split}
\tr\left(A^\top U_{\{j\}}U_{\{j\}}^\top\right) = \sum_{i=1}^d \lambda_i(A) \left(U_{\{j\}}U_{\{j\}}^\top\right)_{ii} \leq \sum_{i=1}^j \lambda_i(A).
\end{split}
\end{equation*}
\quad $\square$

The following lemma characterizes the square-root factorization perturbation. The proof involves Abel's summation identity in Lemmas \ref{lm:Hy-fan-norm} and \ref{lm:inequality}, which is highly non-trivial. 
\begin{Lemma}\label{lm:factorization-lemma}
	Suppose $\hat{M}, M\in \mathbb{R}^{p\times r}$ are two matrices with the same dimension, then there exists an orthogonal matrix $O\in \mathbb{O}_{r}$ such that
	\begin{equation}\label{ineq:D-DO}
	\left\| \hat{M} - MO\right\|_F^2 \leq \frac{\|\hat{M}\hat{M}^\top - MM^\top\|_F^2}{\sigma_r(M)\sigma_r(\hat{M})} \wedge \left(\|\hat{M}\|_F^2 + \|M\|_F^2\right).
	\end{equation}
\end{Lemma}

{\noindent\bf Proof of Lemma \ref{lm:factorization-lemma}.} Suppose $M^\top \hat{M}$ has singular value decomposition: $M^\top \hat{M} = U\Sigma V^\top$, where $U_M, V_M\in \mathbb{O}_r$, $\Sigma\in \mathbb{R}^{r\times r}$. We will show that when $O = UV^\top$ (namely the solution to Wahba's problem), \eqref{ineq:D-DO} holds. 

Define $x_i = \sigma_i(\hat{M}), y_i = \sigma_i(M)$, $z_i = \sigma_i(\hat{M}^\top M)$, by Lemma \ref{lm:Hy-fan-norm}, we know
\begin{equation*}
x_1\geq \cdots \geq x_r \geq 0, \quad y_1\geq \cdots \geq y_r \geq 0,\quad z_1\geq \cdots \geq z_r \geq 0,
\end{equation*}
and $\sum_{i=1}^s z_i \leq \sum_{i=1}^s x_iy_i$ for all $1\leq s \leq r$. Then by both inequalities of Lemma \ref{lm:inequality}, 
\begin{equation*}
\begin{split}
& \sum_{i=1}^r (x_i^4 + y_i^4 - 2z_i^2) - \sum_{i=1}^r\left(x_i^2+y_i^2-2z_i\right)x_ry_r\\
\geq & 2\sum_{i=1}^r (x_i^2y_i^2 - z_i^2) - 2\sum_{i=1}^r (x_iy_i - z_i)x_ry_r \geq 0.
\end{split}
\end{equation*}
On the other hand,
\begin{equation*}
\begin{split}
& \left\|\hat{M} - MO\right\|_F^2 = \tr\left(\hat{M}\hat{M}^\top + MM^\top - \hat{M} O^\top M^\top - M O \hat{M}^\top \right)\\
= & \|\hat{M}\|_F^2 + \|M\|_F^2 - 2\tr(O^\top M^\top \hat{M}) = \|\hat{M}\|_F^2 + \|M\|_F^2 - 2\tr(V U^\top U \Sigma V)\\
= & \sum_{i=1}^r \left(\sigma_i^2(\hat{M}) + \sigma_i^2(M) - 2 \sigma_i(M^\top\hat{M})\right) = \sum_{i=1}^r(x_i^2+y_i^2-2z_i);
\end{split}
\end{equation*}
\begin{equation*}
\begin{split}
& \left\|\hat{M}\hat{M}^\top - MM^\top \right\|_F^2 \\
= & \tr\left(\hat{M}\hat{M}^\top\hat{M}\hat{M}^\top + MM^\top MM^\top - \hat{M}\hat{M}^\top MM^\top - MM^\top \hat{M}\hat{M}^\top \right)\\
= & \|\hat{M}\hat{M}^\top\|_F^2 + \|MM^\top\|_F^2 - 2\|M^\top \hat{M}\|_F^2 \\
= & \sum_{i=1}^r \left(\sigma_i^4(\hat{M}) + \sigma_i^4(M) - 2 \sigma_i^2(M^\top \hat{M}) \right) = \sum_{i=1}^r(x_i^4 + y_i^4 - 2z_i^2),
\end{split}
\end{equation*}
which means
\begin{equation*}
\begin{split}
& \left\|\hat{M}-MO\right\|_F^2\sigma_{r}(M)\sigma_r\hat{M} = x_ry_r\sum_{i=1}^r (x_i^2+y_i^2 - 2z_i)\leq \sum_{i=1}^r \left(x_i^4+y_i^4 - 2z_i^2\right)\\
\leq & \left\|\hat{M}\hat{M}^\top - MM^\top\right\|_F^2,
\end{split}
\end{equation*}
\begin{equation*}
\min_{O\in \mathbb{O}_r} \left\| \hat{M} - MO\right\|_F^2 \leq \frac{\|\hat{M}\hat{M}^\top - MM^\top\|_F^2}{\sigma_r(M)\sigma_r(\hat{M})}.
\end{equation*}
In addition, 
\begin{equation*}
\min_O \left\|\hat{M} - MO\right\|_F^2 \leq \frac{1}{2}\left(\left\|\hat{M} - IM\right\|_F^2 + \left\|\hat{M} + IM\right\|_F^2\right) = \|\hat{M}\|_F^2 + \|M\|_F^2.
\end{equation*}
Therefore, we have finished the proof of this lemma.
\quad $\square$

\begin{Lemma}\label{lm:Hy-fan-norm}
	Suppose $M, \hat{M} \in \mathbb{R}^{p\times r}$ are two matrices of the same dimensions, we have the following inequality for Ky Fan $s$-norm of $M^\top\hat{M}$ \citep{fan1950theorem} for any $s\geq 1$,
	\begin{equation*}
	\|M^\top \hat{M}\|_{k_s} = \sum_{i=1}^s \sigma_i(M^\top \hat{M}) \leq \sum_{i=1}^s \sigma_i(M)\sigma_i(\hat{M}).
	\end{equation*}
\end{Lemma}

{\bf\noindent Proof of Lemma \ref{lm:Hy-fan-norm}.} We first note the following property for Ky Fan norm \citep{fan1950theorem},
$$\|X\|_{k_s} = \sum_{i=1}^s \sigma_i(X) = \max_{\substack{U\in \mathbb{O}_{p, s}\\V\in \mathbb{O}_{r, s}}} \tr(U^\top XV).$$
Let $\hat{M} = U_{\hat{M}}\Sigma_{\hat{M}}V_{\hat{M}}^\top$ be the singular value decomposition, then $(\Sigma_{\hat{M}})_{ii} = \sigma_i(\hat{M})$. Now for any $U, V \in \mathbb{O}_{r, s}$,
\begin{equation*}
\begin{split}
\tr\left(U^\top M^\top \hat{M} V\right) = & \tr\left(U^\top M^\top U_{\hat{M}}\Sigma_{\hat{M}}V_{\hat{M}}^\top V\right) = \tr\left(V_{\hat{M}}^\top V U^\top M^\top U_{\hat{M}}\Sigma_{\hat{M}} \right)\\
= & \sum_{i=1}^r (\Sigma_{\hat{M}})_{ii} \left(U_{\hat{M}}^\top M U V^\top V_{\hat{M}}\right)_{ii} =  \sum_{i=1}^r \sigma_i(\hat{M}) \left(U_{\hat{M}}^\top M U V^\top V_{\hat{M}}\right)_{ii}\\
= & \sum_{i=1}^r\left\{\left(\sigma_i(\hat{M}) - \sigma_{i+1}(\hat{M})\right) \sum_{j=1}^i \left(U_{\hat{M}}^\top M U V^\top V_{\hat{M}}\right)_{jj}\right\},
\end{split}
\end{equation*}
where is last equality is due to the Abel's summation formula\footnote{\url{https://en.wikipedia.org/wiki/Summation_by_parts}}. Note that $U_{\hat{M}}^\top M U V^\top V_{\hat{M}}$ is a $s$-by-$s$ projection of $M$, so it has smaller Ky Fan norms than $M$. Then
\begin{equation*}
\begin{split}
\text{when } i\leq s, \quad & \sum_{j=1}^{i}\left(U_{\hat{M}}^\top M V^\top U  V_{\hat{M}}\right)_{jj} = \sum_{j=1}^i e_j^\top U_{\hat{M}}^\top M U V^\top V_{\hat{M}} e_j \\
\leq & \| U_{\hat{M}}^\top M U V^\top V_{\hat{M}}\|_{k_i} \leq \sum_{j=1}^i\sigma_j(M);\\
\end{split}
\end{equation*}
\begin{equation*}
\begin{split}
& \text{when } i>s, \quad \sum_{j=1}^{i}\left(U_{\hat{M}}^\top M U V^\top V_{\hat{M}}\right)_{jj} \leq \|U_{\hat{M}}^\top M U V^\top V_{\hat{M}}\| \leq \sum_{j=1}^s \sigma_j(M).
\end{split}
\end{equation*}
Thus,
\begin{equation*}
\begin{split}
& \tr\left(U^\top M^\top \hat{M}V\right) \leq  \sum_{i=1}^s\left\{\left(\sigma_i(\hat{M}) - \sigma_{i+1}(\hat{M})\right) \sum_{j=1}^i \left(U_{\hat{M}}^\top M U V^\top V_{\hat{M}}\right)_{jj}\right\}\\
& + \sum_{i=s+1}^r\left\{\left(\sigma_i(\hat{M}) - \sigma_{i+1}(\hat{M})\right) \sum_{j=1}^i \left(U_{\hat{M}}^\top M U V^\top V_{\hat{M}}\right)_{jj}\right\}\\
\leq & \sum_{i=1}^s\left\{\left(\sigma_i(\hat{M}) - \sigma_{i+1}(\hat{M})\right) \sum_{j=1}^i \sigma_j(M)\right\}\\ 
& + \sum_{i=s+1}^r\left\{\left(\sigma_i(\hat{M}) - \sigma_{i+1}(\hat{M})\right) \sum_{j=1}^s \sigma_j(M) \right\}\\ 
= & \sum_{i=1}^s \sigma_i(\hat{M})\sigma_i(M),
\end{split}
\end{equation*}
since $U$ and $V$ are arbitrarily chosen from $\mathbb{O}_{r, s}$, we have finished the proof for this lemma. \quad $\square$

\begin{Lemma}\label{lm:inequality}
	Suppose $\{x_i\}_{i=1}^r, \{y_i\}_{i=1}^r, \{z_i\}_{i=1}^r$ are three sequences of non-negative values satisfying
	$$x_1\geq \cdots\geq x_r\geq 0, \quad y_1\geq \cdots \geq y_r \geq 0, \quad z_1\geq \cdots \geq z_r\geq 0;$$ 
	$$\forall 1\leq s\leq r, \quad \sum_{i=1}^s x_iy_i \geq \sum_{i=1}^s z_i. $$
	This means $x_1y_1\geq z_1$, $x_1y_1+x_2y_2\geq z_1+z_2$, but not necessarily $x_2y_2\geq z_2$. Then, we must have the following two inequalities,
	\begin{equation*}
	\begin{split}
	\sum_{i=1}^r \left(x_i^2y_i^2 - z_i^2\right) - \sum_{i=1}^r (x_iy_i - z_i) x_ry_r \geq 0.
	\end{split}
	\end{equation*}
	\begin{equation*}
	x_i^4 + y_i^4 - (x_i^2 + y_i^2)x_ry_r - 2x_i^2y_i^2 + 2x_iy_ix_ry_r \geq 0,\quad \forall 1\leq i \leq r.
	\end{equation*}
\end{Lemma}

{\noindent\bf Proof of Lemma \ref{lm:inequality}.} The key to the first inequality is via Abel's summation formula. First, 
\begin{equation*}
\begin{split}
& \sum_{i=1}^r (x_i^2y_i^2 - z_i^2) + \sum_{i=1}^r (x_iy_i-z_i)x_ry_r - \left\{\sum_{i=1}^r (x_i^2y_i^2 - z_i^2) -  \sum_{i=1}^r (x_iy_i-z_i)x_iy_i\right\}\\
= & \sum_{i=1}^r (x_iy_i - z_i) (x_iy_i - x_ry_r) = \sum_{i=1}^{r-1} \left\{(x_iy_i - x_{i+1}y_{i+1})\sum_{j=1}^i (x_jy_j-z_j)\right\} \geq 0.
\end{split}
\end{equation*}
If we let $x_{r+1} = y_{r+1} = z_{r+1} = 0$, then
\begin{equation*}
\begin{split}
&  \left\{\sum_{i=1}^r (x_i^2y_i^2 - z_i^2) -  \sum_{i=1}^r (x_iy_i-z_i)x_iy_i\right\} = \sum_{i=1}^r z_i(x_iy_i - z_i)\\
= & \sum_{i=1}^r (z_i - z_{i+1}) \sum_{j=1}^i (x_iy_i - z_i) \geq 0.
\end{split}
\end{equation*}
By combining the two inequalities above, we have finished the proof for the first part. In addition, by some algebraic calculation we can show
\begin{equation*}
\begin{split}
& x_i^4 + y_i^4 - (x_i^2 + y_i^2)x_ry_r - 2\left(x_i^2y_i^2 - x_iy_ix_ry_r\right)\\
= & x_i^4 + y_i^4 - (x_i^2 + y_i^2)x_iy_i - 2\left(x_i^2y_i^2 - x_i^2y_i^2\right) + (x_i^2+y_i^2 - 2x_iy_i)(x_iy_i - x_ry_r)\\
= & x_i^4 + y_i^4 - x_i^3y_i - x_iy_i^3 + (x_i-y_i)^2(x_iy_i - x_ry_r)\\
= & (x_i-y_i)^2(x_i^2+x_iy_i + y_i^2) + (x_i-y_i)^2 (x_iy_i-x_ry_r) \geq 0.
\end{split}
\end{equation*} 
Therefore we have finished the proof for this lemma. \quad $\square$

\end{document}